\pdfoutput=1

\documentclass[fleqn,usenatbib,useAMS]{mnras}

\usepackage{graphicx}	
\usepackage{amsmath}	
\usepackage{multicol}        
\usepackage{bm}		
\usepackage{pdflscape}	
\usepackage{latexsym}
\usepackage[authoryear]{natbib}
\usepackage{bm,nicefrac}

\usepackage[T1]{fontenc}
\usepackage{ae,aecompl}

\usepackage{newtxtext,newtxmath}

\def\deg{\ifmmode^\circ\else$^\circ$\fi}

\def\Q{\ifmmode\mathcal{Q}\else$\mathcal{Q}$\fi}
\def\Mach{\ifmmode\mathcal{M}\else$\mathcal{M}$\fi}

\title[Probing magnetic fields toward G351]
{Unveiling an Hourglass-Shaped Magnetic Field toward IRDC G351.77–0.53}
\author[O.~R. Jadhav et al.]{O.~R. Jadhav$^{1,2}$\thanks{E-mail: omkar@prl.res.in},
L.~K. Dewangan$^{1}$, I.~I. Zinchenko$^{3}$,
Thushara G.~S. Pillai$^{4}$,
\newauthor
Patricio Sanhueza$^{5}$,
A.~K. Maity$^{1}$,
Ram~K. Yadav$^{6}$, and
Saurabh Sharma$^{7}$
\\\\
$^{1}$Physical Research Laboratory, Navrangpura, Ahmedabad - 380 009, India.\\
$^{2}$Indian Institute of Technology Gandhinagar Palaj, Gandhinagar 382355, India.\\ 
$^{3}$Institute of Applied Physics of the Russian Academy of Sciences, 46 Ulyanov st., Nizhny Novgorod 603950, Russia.\\
$^{4}$Haystack Observatory, Massachusetts Institute of Technology, 99 Millstone Road, Westford, MA 01886, USA.\\
$^{5}$Department of Astronomy, School of Science, The University of Tokyo, 7-3-1 Hongo, Bunkyo, Tokyo 113-0033, Japan.\\
$^{6}$National Astronomical Research Institute of Thailand (Public Organization), 260 Moo 4, T. Donkaew, A. Maerim, Chiangmai 50180, Thailand.\\
$^{7}$Aryabhatta Research Institute of Observational Sciences, Manora Peak, Nainital 263002, India.\\
}

\begin{document}

\date{ }

\pagerange{\pageref{firstpage}--\pageref{lastpage}} \pubyear{2023}

\maketitle

\label{firstpage}

\begin{abstract}

We present the SOFIA/HAWC+ 214 $\mu$m polarimetric observations toward the infrared dark cloud G351.77-0.53 (hereafter G351), complemented by existing multi-wavelength data sets. Infrared excess from the embedded sources indicate ongoing star formation activity in the cloud. The G351 cloud hosts two prominent star-forming clumps, i.e., c1 and c2. The plane-of-the-sky magnetic field lines from {\it Planck} observations are predominantly oriented perpendicular to the filament's major axis. Magnetic field orientations from SOFIA/HAWC+ 214 $\mu$m observations reveal distinct hourglass-shaped field configuration toward c1, while the field lines remain perpendicular to the rest of the filament. Using the Davis-Chandrasekhar-Fermi method, we estimate a mean plane-of-the-sky magnetic field strength of $\sim$147 $\pm$ 60 $\mu$G in the G351 filament, with values reaching $\sim$0.8 mG toward c1. The mass-to-flux ratio analysis indicates that the filament is magnetically transcritical, where the gravitational and magnetic field energies are comparable. 
The hourglass-shaped magnetic field observed toward c1 could result from magnetically regulated gravitational collapse, the alignment of converging sub-filaments with the magnetic field, or a combination of both processes. The energy budget analysis further indicates that magnetic fields play an important role in governing the cloud's gas dynamics, followed by contributions from turbulence and gravity.

\end{abstract}
%
\section{Introduction}
\label{sec:intro}

Filaments are prevalent structures within the interstellar medium (ISM) and are linked to both low- and high-mass star formation activities \citep{Arzoumanian_2011, Andre_2014}. These filamentary structures harbor dense cores that eventually form stars. Clouds or filamentary structures that are particularly dense and cold, and appear in absorption against the bright mid-infrared (MIR) Galactic background, are referred to as infrared (IR) dark clouds (IRDCs). Owing to their high column densities ($N(\rm H_2)$) and low temperatures, IRDCs are considered to trace the earliest stages of star formation \citep{Rathborne_2006,Chambers_2009,Sanhueza_2012}. These IRDCs harbor dense cores that eventually give rise to massive stars and stellar clusters. The intense radiative and mechanical feedback from these massive stars disrupts the initial conditions that led to their formation.

Several observational studies have revealed ordered magnetic field (B-field) morphologies toward these filamentary structures \citep{Planck_2016, Stephens_2025, Coude_2025}, indicating that B-fields play a dynamically significant role in regulating the physical processes involved in star formation. The B-field morphology is generally perpendicular to high-$N(\mathrm{H}_2)$ ($> 5 \times 10^{21}$ cm$^{-2}$) \citep{Soler_2017} filaments, whereas in low-$N(\mathrm{H}_2)$ ($< 5 \times 10^{21}$ cm$^{-2}$) filaments, the field orientations can be parallel or more randomly distributed . These trends are observed in serveral magnetohydrodynamical (MHD) simulations as well \citep{Nakamura_2008,Soler_2013, Seifred_2015, Chen_2015, Maity_2024}. However, the exact role of B-fields and their interplay with gravity and turbulence, especially in IRDCs still remains poorly understood. In the literature, the detailed study of B-field structure has been carried out toward only a handful of IRDCs, such as G11.11-0.12 \citep{Chen_2023,Ngoc_2023,Truong_2025}, G14.225-0.506 \citep{Santos_2016,Lopez_2020}, G28.34+0.06 \citep{Law_2024, Liu_2024, Hwang_2025}, G34.43+0.24 \citep{Tang_2019,Soam_2019,Pravash_2025}, G47.06+0.26 \citep{Stephens_2022,Jadhav_2025}, and G11.92 \citep{Sanhueza_2025}.
B-fields are thought to influence star formation processes within IRDCs by potentially reducing small-scale fragmentation, aiding the formation of massive stars, and guiding gas flows along preferred directions; however, the observational evidence remain limited \citep{Pillai_2015}. Therefore, studying the B-field structure in filamentary IRDCs is essential for understanding its role and interplay with turbulence and gravity.

G351.77-0.53 (G351 hereafter) is a well-studied massive star-forming region hosting massive protocluster IRAS 17233-3606 at its center \citep[][and references within]{Simon_2006,Leurini_2008,Zapata_2008,Leurini_2011,Ryabukhina_2021,Beuther_2025,Ishihara_2024,Ginsburg_2023}. The region hosts 6.7 GHz Class~II CH$_3$OH masers \citep{Norris_1993,Walsh_1998,Beuther_2017,Beuther_2019} and several outflow activities \citep{Leurini_2009,Leurini_2011b,Leurini_2013,Klaassen_2015,Garrido_2025}. \citet{Leurini_2019} investigated the G351 cloud with Atacama Pathfinder Experiment (APEX) and \textit{Herschel} observations and revealed that the cloud consists of a main filamentary structure surrounded by a network of filaments. The gas kinematics studies shows that the filament has a coherent velocity of $\sim$3 km s$^{-1}$ \citep{Leurini_2011, Leurini_2019}. The distance to this source has been highly debated, with the values ranging from 0.7 to 2.2~kpc being used in previous studies \citep{Leurini_2011, Ryabukhina_2021, Motte_2022}. However, a recent study of the G351 cloud by \citet{Reyes_2024}, based on Gaia DR3 data, estimated a distance of $\sim$2.0 kpc based on the average parallax of the likely members of the cloud. We adopt this distance for our current work. 

Despite these extensive studies, the structure of the B-field and its influence on massive star formation activity within the G351 cloud remain unexplored. Therefore, a comprehensive investigation using polarimetric observations is essential for characterizing the B-field morphology in this region. Moreover, the relative contributions of magnetic, gravitational, and kinetic energies is crucial to assess whether the B-fields govern the cloud's dynamics or if gravity and turbulence play a more dominant role. In this context, our current study focuses on investigating the role of B-fields using the far-infrared (FIR) polarimetric data to estimate the B-field morphology, strength, and energy. Additionally, this study incorporates complementary multi-wavelength datasets spanning from radio to IR wavelengths.

This paper is organized as follows. In Section~\ref{sec:data}, we describe the data selection. In Section~\ref{sec:result}, we present the outcomes of our multi-wavelength data analysis. In Section~\ref{sec:disc}, we discuss the implications of our results. Finally, the findings are summarized and the conclusions are presented in Section~\ref{sec:summary}.

\section{Datasets used}
\label{sec:data}

In this work, we analyzed multi-wavelength data sets toward the G351 cloud, primarily in the area of 0.22$^\circ$ × 0.27$^\circ$ centered at Right Ascension (RA) = 17h26m33s and Declination (Dec) = -36d08m03s.  A summary of all the data sets analyzed in this study is presented below.

\subsection{NIR and MIR data}

We acquired Near-Infrared (NIR) and MIR images at 3.6--8.0 $\mu$m (resolution $\sim$2$''$; plate scale $\sim$0\rlap.{$''$}6) from the {\it Spitzer} Galactic Legacy Infrared Mid-Plane Survey Extraordinaire \citep[GLIMPSE;][]{Benjamin_2003} survey. 
The MIR 12.0~$\mu$m image was obtained from the Unblurred Coadds of the Wide-field Infrared Survey Explorer (WISE) Imaging (unWISE) survey \citep[resolution $\sim$6$''$;][]{Lang_2014}, which provides WISE images with an improved signal-to-noise ratio.

\subsection{FIR data}
The G351 cloud was observed with the Stratospheric Observatory for Infrared Astronomy (SOFIA) telescope using the High-resolution Airborne Wideband Camera Plus (HAWC+) instrument \citep{Harper_2018} at 214 $\mu$m (or Band E). These observations were taken as part of a survey program (PI: Thushara Pillai; Proposal ID: 07$\_$0238). We used the Level 4 data products in this study which have a spatial resolution of $\sim$18\rlap.{$''$}2 and a pixel scale of $\sim$4\rlap.{$''$}55.

\subsection{Sub-millimeter data}
The 870 $\mu$m emission data (resolution $\sim$18\rlap.{$''$}2) were taken from the Atacama Pathfinder Experiment (APEX) Telescope Large Area Survey of the GALaxy  \citep[ATLASGAL;][]{Schuller_2009}. 
The catalog presented by \citet{Urquhart_2017} provides details for roughly 8,000 dense clumps in the Galactic disc based on the ATLASGAL data. They also provided the velocities, distances, luminosities, and masses of clumps \citep[see][for more details]{Urquhart_2017}. Using this catalog, we identified the positions of dense clumps associated with the G351 cloud.
To infer the POS B-field orientation toward the cloud at larger scales ($\sim$10 pc) we utilized the 353 GHz polarization maps obtained from the Planck Legacy archive observed as part of the Planck mission. The data provides the $I$, $Q$, and $U$ polarization maps observed using the High Frequency Instrument (HFI) \citep{Lamarre_2015} on the Planck satellite, with a spatial resolution of 5$'$ \citep{Planck_2015}.

\subsection{Millimeter data}

The C$^{18}$O(2--1) molecular line data used in this study are the same as those analyzed by \citet{Ryabukhina_2021}, obtained from observations conducted with the APEX radio telescope between 2010 and 2017 (projects O-085.F-9323, O-086.F-9316, O-097.F-9303, O-098.F-9306) using SHeFI receivers. The resolution of the data is $\sim$32$''$. The data are converted to the main beam temperature scale using the main beam efficiencies ($\eta_{\rm mb}$) presented on the APEX web site. The antenna temperature is converted into main beam temperature using the formula $T_{\rm mb}$ = $T_{\rm A}$/$\eta_{mb}$, where $\eta_{\rm mb}$ = 0.75. Further details of the APEX C$^{18}$O(2--1) observations and data processing can be found in \citet{Ryabukhina_2021}.

\subsection{Radio data}
We have used the South African Radio Astronomy Observatory (SARAO) MeerKAT Galactic Plane Survey (SMGPS) 1.3 GHz continuum emission data \citep[resolution $\sim$8$''$; sensitivity $\sim$22 $\mu$Jy beam$^{-1}$;][]{Goedhart_2024}.

\section{Results}
\label{sec:result}

\subsection{Physical environment toward the G351 IRDC}

\begin{figure*}
\centering
\includegraphics[width=\textwidth]{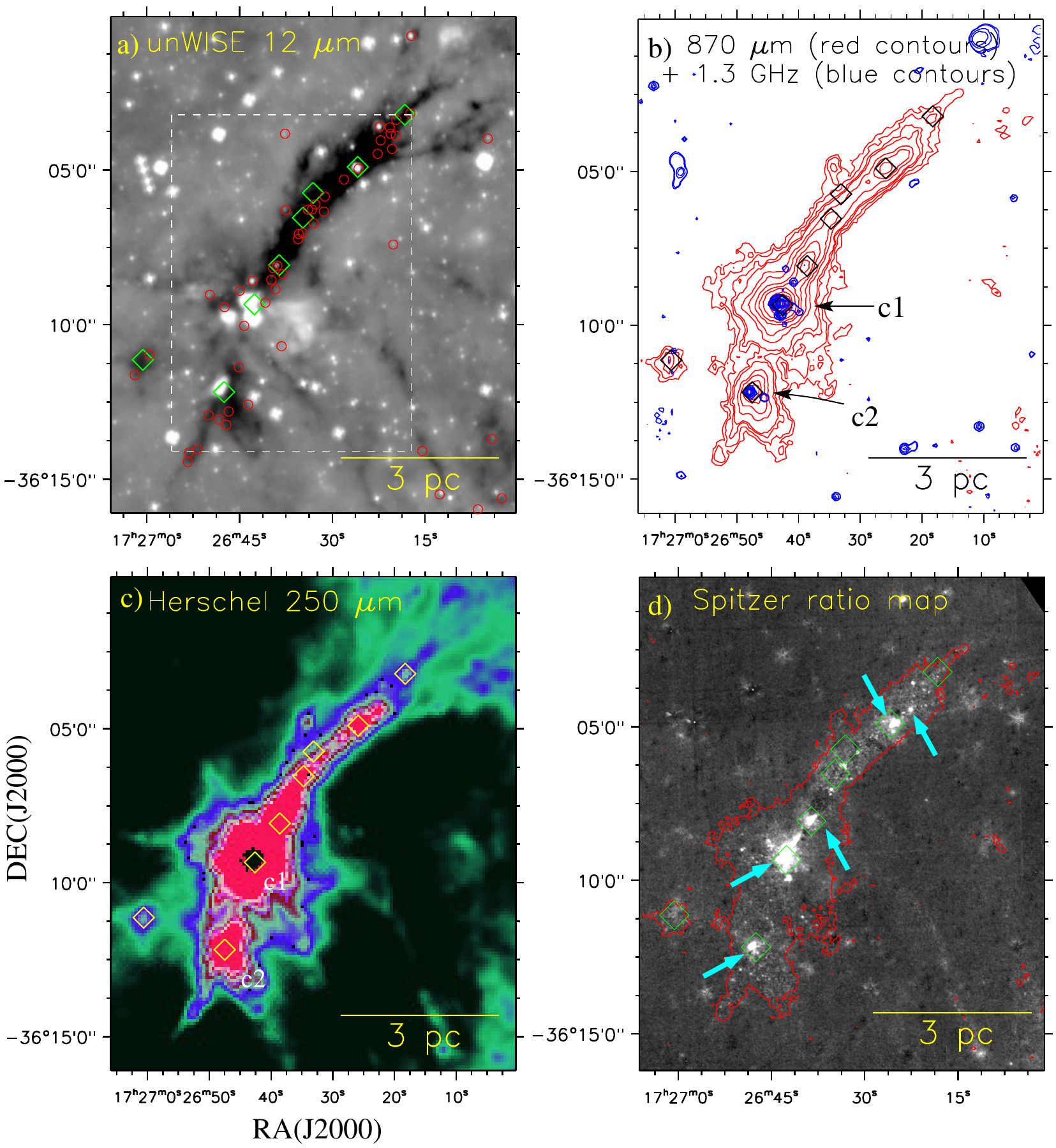}
\caption{a) unWISE 12.0~$\mu$m image of the G351 cloud overlaid with the positions of YSOs (red circles) identified using [5.8]-[4.5] vs [3.6]-[4.5] CCD (see text for more details). The white dashed box highlights the area shown in Figure~\ref{fg3}a--~\ref{fg3}d. b) An overlay of SMGPS 1.3 GHz continuum emission contours (in blue) on ATLASGAL 870 $\mu$m contours (in red). The red contour levels are 0.14, 0.24, 0.38, 0.48, 0.94, 1.41, 2.35, 3.30, 4.71, 9.42, 14.12, 18.83, 23.54, 37.66, and 42.37 mJy beam$^{-1}$. The blue contour levels are 0.44, 0.88, 2.6, 4.4, 6.2, 8.8, 18, 26, 35, 44, 70, and 79 mJy beam$^{-1}$. c) \textit{Herschel} 250 $\mu$m image of the G351 cloud. d) \textit{Spitzer} 4.5 $\mu$m/3.6 $\mu$m ratio map of the G351 cloud overlaid with the 870 $\mu$m emission contour level at 0.14 mJy beam $^{-1}$. The cyan arrows are used to highlight the bright regions where star formation activity is prominent. The diamonds show the positions of ATLASGAL clumps. The scale bar corresponds to 3 pc at a distance of 2.0 kpc. }
\label{fg1}
\end{figure*}

Figure~\ref{fg1}a presents the unWISE 12.0~$\mu$m image of the G351 cloud, overlaid with diamonds marking the positions of the ATLASGAL clumps. The image reveals the filamentary structure of G351 seen in absorption against the MIR background. Most of the ATLASGAL clumps are distributed toward the northern region of the filament. We also identified Class~I young stellar objects (YSOs) toward the cloud using the [4.5]$-$[5.8] versus [3.6]$-$[4.5] color–color diagram (CCD) \citep{Hartmann_2005,Getman_2007}. Although these plots are not shown in this paper, further details can be found in \citet{Dewangan_2015}. A total of 53 Class~I YSOs are identified using this method, and their locations are marked with red circles in Figure~\ref{fg1}a. Most of these Class~I YSOs are seen toward the filament, confirming ongoing star formation activity within the filamentary cloud. Figure~\ref{fg1}b shows the ATLASGAL 870 $\mu$m contours overlaid with SMGPS 1.3 GHz continuum emission contours, revealing substructures within the cloud.
%
%
The SMGPS 1.3 GHz continuum emission is concentrated only toward two clumps (c1 and c2), while the rest of the cloud shows little to no radio emission. The absence of radio emission toward most of the filament indicates that the filament is in an early stage of star formation.
To further explore the star formation activity, we generated \textit{Spitzer} 4.5~$\mu$m/3.6~$\mu$m ratio map. Figure~\ref{fg1}d displays the resulting ratio map, showing several bright areas (see arrows in Figure~\ref{fg1}d). These bright areas are dominated by the 4.5~$\mu$m emission and are consistent with the locations of the dust clumps. This excess 4.5~$\mu$m emission could be due to the ionized emission at 4.05 $\mu$m or the H$_2$ emission at 4.69 $\mu$m. Given the presence of Class~I protostars towards these bright areas, the emissions likely correspond to H$_2$ features, while the darker areas trace polycyclic aromatic hydrocarbon (3.3 $\mu$m PAH emission in {\it Spitzer} 3.6 $\mu$m band), indicating the presence of warm dust.

\subsection{POS B-field orientations from the SOFIA/HAWC+ and Planck}
\label{sec:B-field}

\begin{figure*}
	\centering
	\includegraphics[width=18cm]{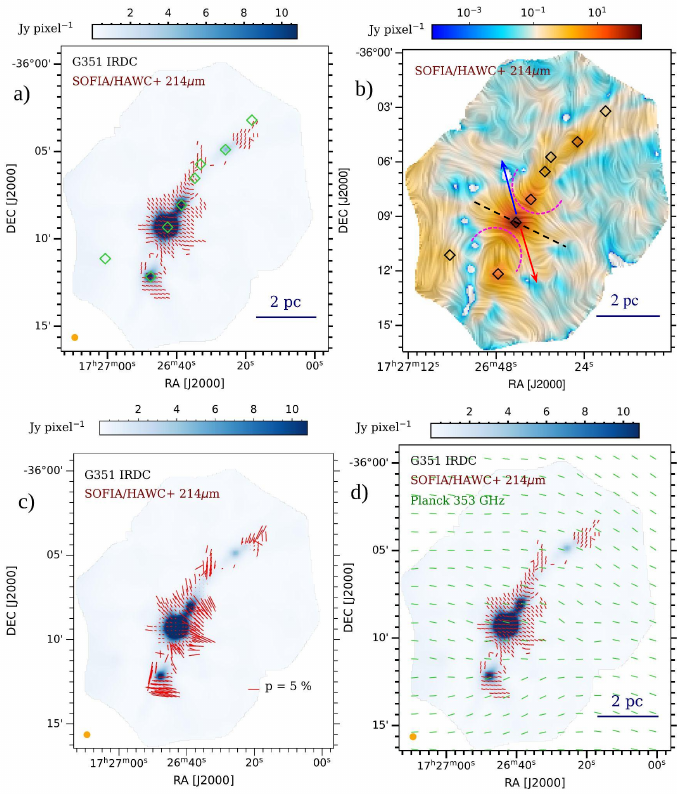}
	\caption{a) SOFIA/HAWC+ 214 $\mu$m image of the G351 overlaid with the red segments showing the direction of the B-field derived using the SOFIA/HAWC+ 214 $\mu$m polarization observations. b) The LIC map showing the B-field same as Figure~\ref{fg2}a, overlaid on the SOFIA/HAWC+ 214 $\mu$m intensity map. The magenta curves highlight the hourglass-shaped B-field. The red and blue arrows show the direction of outflow toward IRAS 17233--3606. The black dashed line shows the direction of velocity gradient \citep{Klaassen_2015}.
	c) Same as Figure~\ref{fg2}a, with length of the segments proportional to the polarization fraction. A reference segment corresponding to a 5~\% polarization fraction is shown in the bottom-right corner of the panel. d) Same as Figure~\ref{fg2}a, overlaid with B-field segments (in green) inferred using the Planck 353 GHz data. The diamonds show the positions of ATLASGAL clumps. The beam size of the SOFIA/HAWC+ 214~$\mu$m observations ($\sim$18\rlap.{$''$}2) is indicated by a filled orange circle. The scale bar corresponds to 2 pc at a distance of 2.0 kpc.}
	\label{fg2}
\end{figure*}

To infer the POS B-field morphology toward the G351 cloud, we employ the SOFIA/HAWC+ 214 $\mu$m polarization data (resolution $\sim$18\rlap.{$''$}2). The $Q$, and $U$ Stokes maps are used to obtain the polarization angle. The polarization angle is estimated using 
\begin{eqnarray}
\theta=\frac{1}{2}~{\rm arctan2}(U, Q)
\end{eqnarray}
The B-field position angles are obtained by rotating the polarization vectors by 90$^\circ$.  We retained the vectors which satisfy the following criteria: $p/\sigma_p>3$, $I/\sigma_I>100$, and $p<30$ \citep{Zelinski_2022, Li_2022,Tram_2023,Sinjen_2025}. Figure~\ref{fg2}a
shows the overlay of the segments showing the SOFIA/HAWC+ 214~$\mu$m B-field orientations on the corresponding intensity map of the G351 cloud. The B-field lines exhibit an hourglass morphology toward the central massive clump (c1), whereas in the rest of the cloud (including c2), the field orientations are predominantly perpendicular to the major axis of the filament. This behavior is also more clearly illustrated in the Line Integral Convolution (LIC) map of the B-fields presented in Figure~\ref{fg2}b. The LIC map is generated using the {\tt magnetar} package \citep{Soler_2013}. 
In Figure~\ref{fg2}b, the red and blue arrows indicate the direction of the outflow, while the black dashed line shows the direction of the velocity gradient reported by \citet{Klaassen_2015}. Since the study by \citet{Klaassen_2015} was conducted at core scales ($\sim$0.01 pc), a direct and detailed comparison with the B-field morphology traced by the SOFIA/HAWC+ observations in this study is not feasible, and require high-resolution polarimetric observations.
Figure~\ref{fg2}c shows the segments indicating the B-field orientations from SOFIA/HAWC+ 214 $\mu$m data, with their lengths corresponding to the degree of polarization. We also inferred the POS B-field direction using the Planck~353 GHz polarization data (see Figure~\ref{fg2}d). The B-field vectors inferred from the Planck data are predominantly perpendicular to the G351 cloud.

\subsection{Molecular gas kinematics}

To study the gas kinematics toward the G351 cloud, we utilized the  C$^{18}$O(2--1) molecular line observations from APEX telescope. 

\subsubsection{Moment maps and PV diagram}
\begin{figure*}
	\centering
	\includegraphics[width=15cm]{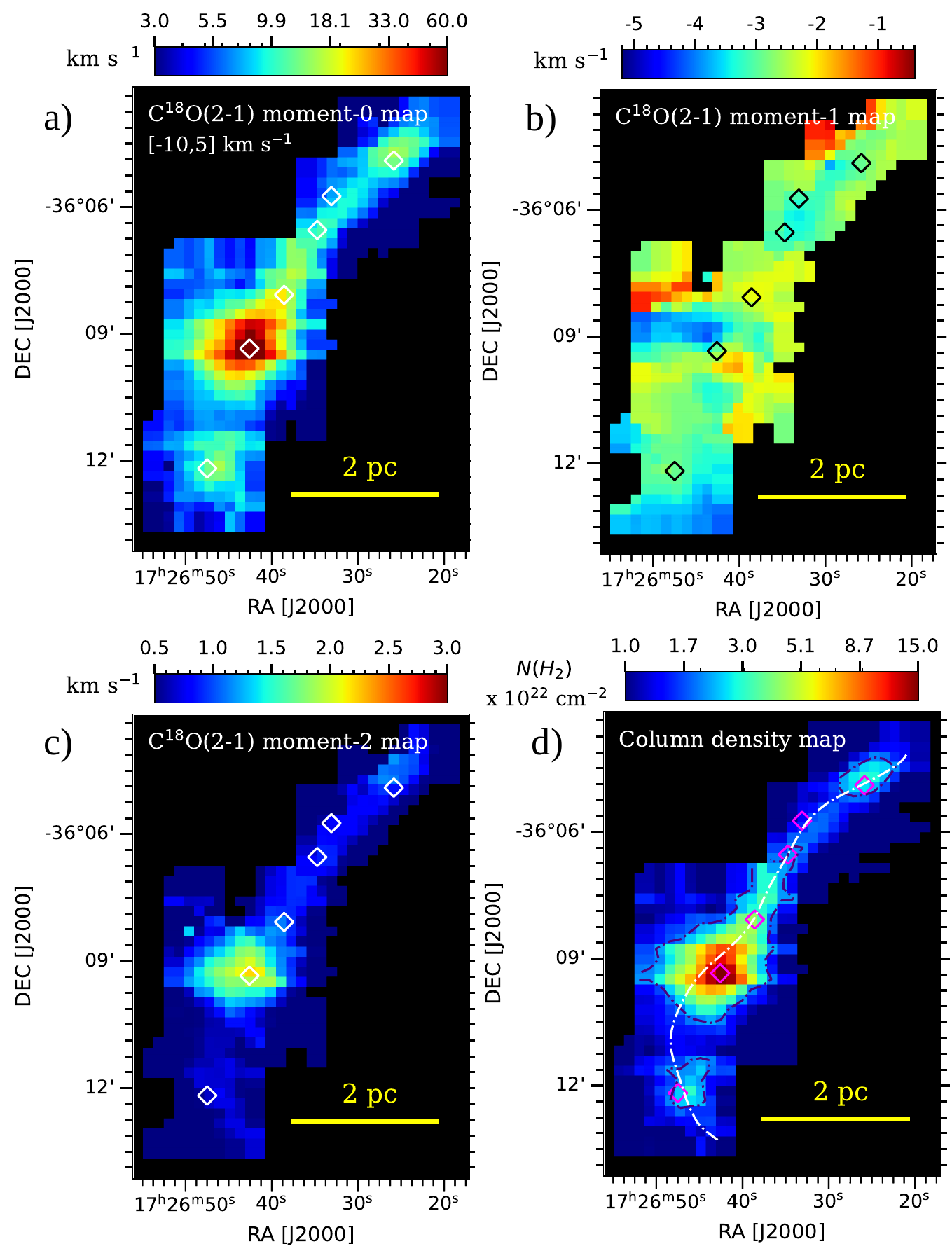}
	\caption{a) C$^{18}$O(2--1) moment-0 map of the G351 cloud integrated over a range of [$-$10, 5] km s $^{-1}$. b) C$^{18}$O(2--1) moment-1 map. c) C$^{18}$O(2--1) moment-2 map. d) $N(\rm H_2)$ map of the G351 cloud. The contour level is 1.6 $\times$ 10$^{22}$ cm$^{-2}$. The dot-dashed line show the spine of the filament identified using the {\tt RadFil} algorithm. The scale bar and symbols are same as Figure~\ref{fg2}. }
	\label{fg3}
\end{figure*}

\begin{figure*}
	\centering
	\includegraphics[width=12cm]{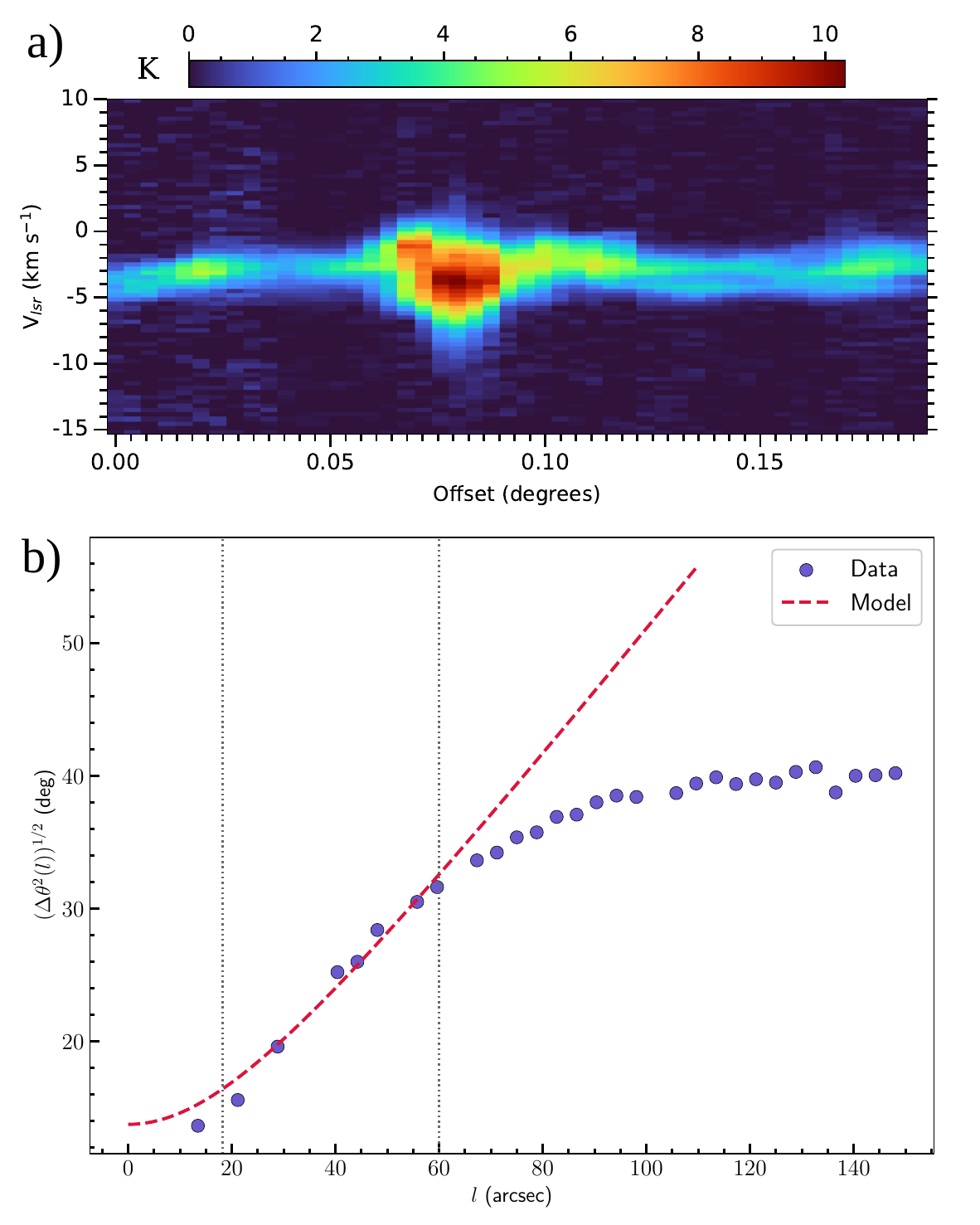}
	\caption{a) PV diagram along the spine of the filamentary cloud shown in Figure~\ref{fg3}d. b) The panel presents the structure function ($D^{1/2}_\theta(\ell)$) for the G351 cloud. The red dashed line shows the linear fit of Equation~\ref{eq:quadratic_func}, while the gray vertical dashed lines mark the lower and upper limits used for the fit.}
	\label{fg4}
\end{figure*}

Figure~\ref{fg3}a shows the C$^{18}$O(2--1) integrated intensity (moment-0) map toward the G351 cloud between [-5, 10] km s$^{-1}$. The moment-0 map retains the filamentary structure of the cloud, with high intensity toward c1. The intensity-weighted mean velocity (moment-1) map is shown in Figure~\ref{fg3}b, which allows us to examine the velocity distribution within the cloud. The map indicates that the entire filament exhibits a coherent velocity of approximately $-$3 km s$^{-1}$. The moment-2 map (see Figure~\ref{fg3}c) allows us to study the velocity dispersion toward the cloud. The moment-2 map shows high velocity dispersion (up to $\sim$2 km s$^{-1}$) toward the c1. To further understand the gas kinematics toward the G351 cloud, we generated the $C^{18}$O(2--1) position-velocity (PV) diagram along the spine of the filament (see Figure~\ref{fg4}). 

\subsubsection{Mass estimation using C$^{18}$O(2--1)}
\label{sec:mass}
We have calculated the gas mass of the G351 cloud using the C$^{18}$O(2--1) emission. 
Assuming the system in local thermodynamic equilibrium (LTE), we can calculate the column density of a linear molecule using the equation \citep{Magnum_2015}  
\begin{eqnarray}
N&=&\frac{3h}{8\pi^3\mu^2S}\frac{Q_\mathrm{rot}}{g_{J}g_{I}g_{K}}
\frac{\mathrm{exp}\left(\frac{E_\mathrm{up}}{k T_\mathrm{ex}}\right)}{\mathrm{exp}\left(\frac{h\nu}{k T_\mathrm{ex}}\right) - 1} \nonumber \\
& &\times\frac{1}{J(T_\mathrm{ex})-J(T_\mathrm{bg})}\int T_\mathrm{r}~dv \label{eq-lte-density},
\end{eqnarray}
where the integral is performed over velocity range [-5, 10] km s$^{-1}$. Here $J(T) = \frac{h\nu/k}{\mathrm{exp}(h\nu/k T)-1}$ and $T_{\rm bg}$ = 2.73~K is the cosmic microwave background temperature. The excitation temperature ($T_\mathrm{ex}$) for the target source is assumed to be 20\thinspace K (mean $T_{\rm d}$ of ATLASGAL clumps toward the filament). The symbol $\mu$ stands for the dipole moment of the molecule. The degeneracy ($g_\mathrm{J})$ is $2J_u+1 = 5$, $g_\mathrm{I}$ is nuclear spin degeneracy, and $g_\mathrm{K}$ is K degeneracy, both are equal to 1 (for C$^{18}$O). The line strength ($S$) is $\frac{J_u}{2J_u+1} = \frac{2}{5}$, where $J_u$ is the rotational quantum number of the upper state (i.e., $J_u = 2$ for the $J = 2-1$ transition). $E_\mathrm{up}$ is the energy of the upper state. We have adopted an approximated formula for the rotational partition function ($Q_\mathrm{rot}$) mentioned in various previous studies \citep[e.g.,][]{Magnum_2015,Yuan_2016}, $Q_\mathrm{rot} = \frac{kT}{hB} + \frac{1}{3}$, where $B$ is the rotational constant of the molecule. Other molecular parameters have been adopted from the Jet Propulsion Laboratory (JPL) Molecular Spectroscopy database and spectral line catalog \citep{Pickett_1998}. 

Using the integrated C$^{18}$O(2--1) emission at [-10, 5]~km s$^{-1}$ in Equation~\ref{eq-lte-density}, the C$^{18}$O column density map is produced. Now considering the column density ratio between H$_2$ and C$^{18}$O \citep[i.e., $\frac{N(\mathrm{H_2})}{N(\mathrm{C^{18}O})}$ = $4.8 \times 10^6$ at a Galactocentric distance of 6.4 kpc;][]{Liu_2013}, we have obtained the $N(\mathrm{H_2})$ map. The resultant column density map is shown is Figure~\ref{fg3}d. We estimate the mass of the cloud using
\begin{eqnarray}
M= \mu_{\rm H_{2}}~m_{\rm H}~ a_{\rm pixel}~\Sigma N(\mathrm H_2),
\label{eq1}
\end{eqnarray}
where $\mu_{\rm H_{2}}$ is the  ${\rm H_2}$ mean molecular weight (assumed to be 2.8), $m_{\rm H}$ is the mass of the hydrogen atom, $a_{\rm pixel}$ is the physical area subtended by 1 pixel in cm$^2$, and $\Sigma N(\mathrm H_2)$ is the integrated column density above contour level of 1.6 $\times$ 10$^{22}$ cm$^{-2}$ (see Figure~\ref{fg3}d). The total mass of the molecular cloud is $\sim$4150 $M_{\odot}$. The uncertainty in the derived mass is dominated by the adopted gas-to-dust ratio and the assumption of LTE, contributing at least a 50\% uncertainty. However, the actual uncertainty may be significantly higher. This estimated mass is roughly half of the previous estimate for this filament \citep{Ryabukhina_2021}, when scaled to the same distance. The difference arises because the previous study used a dust temperature map as $T_{\rm ex}$ to derive the $N(\mathrm{H_2})$ at each pixel, whereas in this work we adopt a constant $T_{\rm ex}$ value. This is due to the saturation of the {\it Herschel} 160, 250, and 350 $\mu$m bands toward the central clump c1. In addition, previous studies adopted a slightly larger integration area and a different C$^{18}$O--H$_2$ conversion factor, which together result in a lower total mass estimate in our study.

\subsection{Estimation of magnetic field strength}

We employ the SOFIA/HAWC+ 214 $\mu$m polarization observations to estimate the POS B-field strength ($B_{\rm POS}$) toward the G351 cloud. The $B_{\rm POS}$ value is estimated using the Davis-Chandrasekhar-Fermi (DCF) method \citep{Chandrasekhar_1953}. However, we use a variant of the DCF method by \citet{Crutcher_2012} to estimate the $B_{\rm POS}$ using, 
\begin{equation}\label{eq1}
B_{\rm POS} =\mathcal{Q}\sqrt{4\pi\rho}~\frac{\sigma_{\rm V_{\rm NT}}}{\delta_\theta} \approx 9.3~\sqrt{n(\rm H_2)}~\frac{\Delta V_{\rm nt}}{\delta_{\theta}}~(\mu G),
\end{equation}
where $Q$ is the correction factor taken as 0.5, $\rho$ is the gas density in g cm$^{-3}$, $\sigma_{\rm V_{\rm NT}}$ is the non-thermal velocity dispersion  in km s$^{-1}$, $\delta_{\theta}$ is the polarization angle dispersion in degrees, $n(\rm H_2)$ is the volume density in cm$^{-3}$, and $\Delta V_{\rm nt}$ is the FWHM of the non-thermal velocity component in km s$^{-1}$. In the following subsections, we derive these parameters for the G351 cloud regions to estimate $B_{\rm POS}$ values.

\subsubsection{Volume density}

Using the $N(\rm H_2)$ map and mass derived in Section~\ref{sec:mass}, the volume density ($n({\rm H_2})$) of the filament can be obtained using the following equation
\begin{eqnarray}
n({\rm H_2}) = \frac{ML}{\pi R^2},
\end{eqnarray}
where $M$ is the mass of the filament, $L$ is the length of the filament, and $R$ is the radius of the filament. The length of the filament is calculated using the \texttt{RadFil}\footnote{https://github.com/catherinezucker/radfil}  algorithm  \citep{Zucker_2018} on the 
$n(\rm H_2)$ map obtained in Section~\ref{sec:mass}. This analysis yields a filament length of approximately 7.8~pc. The \texttt{RadFil} package allows to fit the $n(\rm H_2)$ profile along the spine of the filament with a Plummer-like Function \citep{Andre_2014}. The radius of the filament is adopted to be equal to its flattening radius, as derived from the fitted profile i.e., $\sim$0.55~pc. To generate the volume density map of the G351 cloud, we adopt the relation 
$n(\rm H_2)$ = $N(\rm H_2)/W$, where W is the width of the filament. The width of the filament is taken as $2 \times R_{\mathrm{flat}}$ \citep{Andre_2014}.

\subsubsection{Polarization angle dispersion}
We use the structure function method to estimate the polarization angle dispersion toward G351 \citep{Hildebrand_2009,Houde_2009}. According to this method, the $B_\mathrm{POS}$~can be estimated using Eq.~\ref{eq1} but the polarization angle dispersion $\delta\theta$ is replaced by the structure function of the polarization angles, so-called angular structure function. The angular structure function, $D^{1/2}_\theta$, at an angular scale $\ell$ is estimated as \citep{Hildebrand_2009}:
\begin{equation}
\label{eq:SF}
D^{1/2}_\theta(\ell) \equiv \left< \Delta \theta^2 (\ell) \right> ^{1/2}
= \left\{ \frac{1}{N(\ell)} \sum_{i=1}^{N(\ell)} \Bigl[ \theta(x) - \theta(x+\ell)\Bigr]^2\right\}^{1/2}, 
\end{equation} 
where $\left< ... \right>$ denotes an average, $\Delta\theta(\ell) \equiv\theta(x) - \theta(x+\ell)$ is the polarization angle difference between individual pairs of polarization vectors at position $x$ and $x+\ell$, where $\theta(x)$ and $\theta(x+\ell)$ are the corresponding polarization angles at position $x$ and $x+\ell$, respectively, and $N(\ell)$ is number of pairs of vectors with a displacement of $\ell$. 
At scales $\ell$ much smaller than the scale for a variation of the large-scale B-field structure (typically of $\sim$5$'$, as seen by the \textit{Planck} observations), the total angular dispersion function can be expressed as \citep{Schleuning_1998,Houde_2004,Hildebrand_2009}:
\begin{equation}\label{eq:quadratic_func}
D_\theta(\ell) \simeq b^2 + m^2\ell^2 + \sigma_\mathrm{M}^2(\ell), 
\end{equation}
where $b$ represents the turbulent component of B-fields ($B_\mathrm{t}$), $m\ell$ describes the large-scale structure of B-fields ($B$), and $\sigma_\mathrm{M}(\ell)$ is subject to measurement uncertainties. Also, the ratio between the turbulent and large-scale B-field strength indicates the angular dispersion, given as \citep{Hildebrand_2009}:
\begin{equation}
\delta\theta\simeq \frac{\left<B^2_\mathrm{t}\right>^{1/2}}{B} =\frac{b}{\sqrt{2-b^2}}.
\end{equation}
When $B_\mathrm{t} \ll B $, i.e., $b\ll1$~rad, we can obtain $\delta\theta \sim b/\sqrt{2}$. We calculated the $D^{1/2}_\theta(\ell)$ for G351 using equation~\ref{eq:SF} and fitted with the quadratic function given as equation~\ref{eq:quadratic_func}. The structure function plot for G351 is shown in Figure~\ref{fg4}b. We only fitted the data points between $\sim$18\rlap.{$''$}2 and $\sim$40{$''$}, to avoid fitting large-scale structures.
The lower fitting limit was set to the resolution of the polarization data, whereas the upper limit was chosen based on visual inspection of the structure function, corresponding to the scale just before it reached saturation.
The best-fit parameters for G351 are $b$ = 11.8$^\circ$ $\pm$ 1.5$^\circ$ and $m$ = 0.53$^\circ$ $\pm$ 0.01$^\circ$
The polarization angle dispersion ($\delta_{\theta}$) for the G351 cloud is 8.3$^\circ$ $\pm$ 1.5$^\circ$.
We further note that the angular dispersion analysis adopted here does not explicitly account for the integration of polarized emission along the LOS and within the telescope beam. \citet{Houde_2009} showed that such beam and volume integration effects systematically reduce the observed angular dispersion by averaging over multiple turbulent cells, which in turn leads to an overestimation of the $B_{\rm POS}$ if uncorrected. Incorporating this effect generally results in a lower effective angular dispersion and therefore a reduction of the inferred $B_{\rm POS}$, in some cases by factors of a few, depending on the physical beam size and the turbulent correlation length.

We also utilized the sliding box method presented in \citet{Hwang_2021}, to generate a pixel-by-pixel polarization angle dispersion map. We proceed by defining a box of 20 $\times$ 20 pixels (pixel scale $\sim$4\rlap.{$''$}55), which is approximately 5 times the beam size of the SOFIA/HAWC+ band D observations ($\sim$18\rlap.{$''$}2) centered on the $i$th pixel \citep{Guerra_2021, Ngan_2024}. We estimate the root-mean-squared (RMS) of all the pixel values inside the box. This RMS value corresponds to the polarization angle dispersion at $i^{th}$ pixel. This process is repeated by sliding the box across the entire region to generate the polarization dispersion map. The dispersion value for a given box is computed only if more than 50$\%$ of the pixels within it have valid (non-NaN) values; otherwise, the central pixel is excluded from the calculation. The mean polarization angle dispersion values toward the G351 cloud is 12.5$^\circ$ $\pm$ 5.3$^\circ$. The polarization dispersion values toward G351 using both methods are nearly identical and cannot be distinguished within the error bars.

\subsubsection{Velocity dispersion}

The C$^{18}$O(2--1) molecular line data is employed to estimate  $\Delta V_{\rm nt}$ for the G351 cloud. We obtained $\sigma_{\rm v}$ using the C$^{18}$O(2--1) moment-2 map. The FWHM of the velocity component is calculated as, $\Delta V$ = 2.355 $\times$ $\sigma_{\rm v}$. However, this value also consists the contribution of the thermal component of the velocity. To obtain the non-thermal velocity component we use the equation \citep{Fuller_1992, Kauffmann_2013}
\begin{equation}
{\Delta V_{\rm nt}}^2 = {\Delta V}^2 - {\rm 8~ln2}\frac{kT}{m_{\rm C^{18}O}} ,
\end{equation}
where $k$ is the Boltzmann constant, $T$ is the gas temperature (20 K), and $m_{\rm C^{18}O}$ is the mass of C$^{18}$O molecule which is 30 amu. Figure~\ref{fg3}c shows the C$^{18}$O(2--1) moment-2 map ($\Delta V$) toward the G351 cloud. The mean $\Delta V_{\rm nt}$ value toward G351 is 1.47 $\pm$ 0.41 km s$^{-1}$.

\subsubsection{Magnetic field strength and mass-to-flux ratio}
\label{sec:magnetic strength}

To derive the $B_{\mathrm{POS}}$ map toward the G351 cloud, $n(\mathrm{H_2})$, $\delta\theta$, and $\Delta V$ maps to a common angular resolution ($\sim$32$''$) and pixel scale ($\sim$14\rlap.{$''$}32), same as the C$^{18}$O(2--1) molecular line data. By substituting the $n(\mathrm{H_2})$, $\delta\theta$, and $\Delta V$ maps into equation~\ref{eq1}, we derived the $B_{\mathrm{POS}}$ map toward the G351 cloud (Figure~\ref{fg5}a). The $B_{\mathrm{POS}}$ values reach up to $\sim$0.8 mG toward c1. The mean $B_{\mathrm{POS}}$ obtained from the map is $\sim$170 $\pm$ 127 $\mu$G across the filament. Using the structure function (SF) method, we derived a mean $B_{\mathrm{POS}}$ of $\sim$147 $\pm$ 60 $\mu$G.

To infer the relative importance of B-fields with respect to gravity, we use a parameter known as the mass-to-flux ratio, $\lambda$, which is estimated using the relation given by \citep{Crutcher_2004}

\begin{equation}
\lambda_{\rm obs}=\frac{(M/\Phi)_{\rm observed}}{(M/\Phi)_{\rm critical}}=7.6\times10^{-21}\frac{N(\rm H_2)}{B_{\rm tot}},
\end{equation}
where $B_{\rm tot}$ is the total B-field strength in $\mu$G which is given as, $B_{\rm tot}$ = 1.3 $\times$ $B_{\rm POS}$ \citep{Crutcher_2004}. If $\lambda>1$ signifies the dominance of gravity over the B-field, hence referred to as magnetically supercritical. Conversely, if $\lambda<1$, it denotes that the B-field is dominant over gravity, characterizing those areas as magnetically subcritical. The mass-to-flux ratio toward the G351 cloud is $\sim$0.9 $\pm$ 0.6, characterizing the cloud as magnetically transcritical. Although the cloud is magnetically transcritical, the mass-to-flux ratio map shows $\lambda>1$ values near the c1, indicating a relative dominance of gravity over the B-field.

\subsection{Energy Balance}
\label{sec:energy}

Star formation activity in a cloud is driven by a complex interplay of gravity, turbulence, and B-fields. Therefore, quantifying the relative contributions of these forces is essential to understand their roles in the physical processes governing star formation. Below present the virial theorem including these energies.
\begin{eqnarray}
\frac{1}{2} \text{\"{I}}~=2E_{\rm K} + E_{\rm B} + E_{\rm G}
\end{eqnarray}
Assuming a cylindrical geometry for the filament, the gravitational energy is given by \citep{Fiege_2000}

\begin{eqnarray}\label{Eg}
E_{\rm G} = \frac{-GM^2}{L},
\end{eqnarray}
where $M$ is the mass of the filament ($\sim$4150 $M_{\odot}$), and $L$ is the length of the filament ($\sim$7.8 pc).
The kinetic energy is given by \citep{Fiege_2000}
\begin{eqnarray}\label{Ek}
E_{\rm K} = M\sigma_V^2,
\end{eqnarray}
where $\sigma_V$ is the total velocity dispersion.
The magnetic energy is given by \citep{Crutcher_2012}
\begin{eqnarray}\label{Eb}
E_{\rm B} = \frac{B^2V}{8\pi},
\end{eqnarray}
where $B$ is the POS strength of B-field, $V$ is the volume of the cylinder. The $E_{\rm G}$, $E_{\rm K}$, and $E_{\rm B}$ values for G351 are 1.88 $\pm$ 1.88 $\times$ 10$^{47}$ ergs, 1.79 $\pm$ 1.04 $\times$ 10$^{47}$ ergs, and 3.20 $\pm$ 2.24 $\times$ 10$^{47}$ ergs, respectively.
Taking into account the uncertainties in the mass estimates compared to the previous studies, discussed in Section~\ref{sec:mass}, the derived gravitational, kinetic, and magnetic energies could be underestimated by up to factors of four, two, and two, respectively. Within the associated uncertainties, the results indicate that the B-field, gravity, and turbulence contribute comparably to the overall energy budget of the cloud.

\section{Discussion}
\label{sec:disc}

The SOFIA/HAWC+ 214 $\mu$m polarization observations revealed hourglass-shaped POS B-field toward clump c1 in the G351 cloud (Section~\ref{sec:B-field}). This morphology can be interpreted in two ways: either as a consequence of gravitational contraction, or as the result of converging sub-filaments aligned with the B-field, channeling material into the dense central hub. Both scenarios emphasize the role of B-fields in regulating mass assembly. In the following sections (i.e., Section~\ref{sec:disc1} and \ref{sec:disc2}), we examine these two possibilities in detail.

\subsection{Hourglass-shaped B-field due to magnetically regulated collapse}
\label{sec:disc1}

\begin{figure*}
	\centering
	\includegraphics[width=\textwidth]{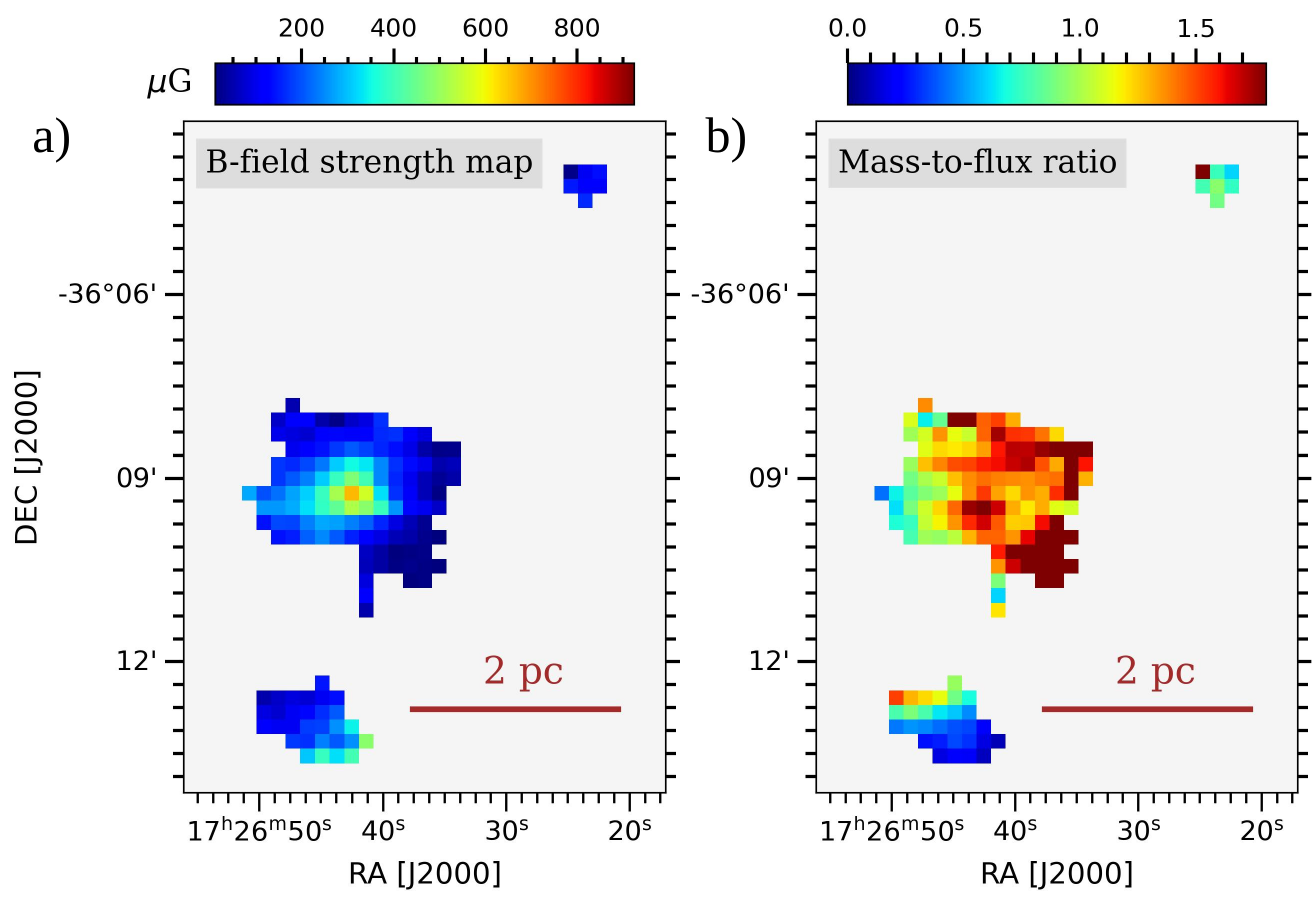}
	\caption{a) B-field strength map of the G351 cloud. b) Mass-to-flux ratio map of the G351 cloud. The scalebar in each panel is same as Figure~\ref{fg2}. }
	\label{fg5}
\end{figure*}

The POS B-field inferred from the Planck observations is predominantly perpendicular to the major axis of the G351 filament, consistent with previous dust polarization studies of filamentary clouds \citep[e.g.,][]{Planck_2016,Soler_2017}. 
The high-resolution SOFIA/HAWC+ 214~$\mu$m observations reveal an hourglass morphology of the B-field
toward the massive star cluster associated with c1/IRAS 17233-3606 (see Figure~\ref{fg2}a). The B-field lines in the rest of the filament are perpendicular to the major axis of the filament, consistent with the Planck observations, except the northernmost and southernmost regions of the filament, where the B-field orientations are aligned parallel to the spine of the filament. Using the DCF method, the $B_{\rm POS}$ toward the G351 cloud is estimated to be $\sim$147 $\pm$ 60 $\mu$G, while the $B_{\rm POS}$ map (see Figure~\ref{fg5}a) reveals values up to 0.8 mG near c1. The B-field energy and strength estimates together indicate that the B-fields are dynamically important in the G351 cloud (see Section~\ref{sec:energy}).
Through numerical simulation, \citet{Nakamura_2008} discussed the evolution of a magnetized filamentary cloud. In this scenario, the initial system is dominated by the B-field, followed by turbulence and gravity. The subsequent evolution depends on the relative contributions of turbulence and gravitational forces. As discussed in \citet{Nakamura_2008}, when turbulence is insufficient to counteract gravitational settling along the B-field lines (see also \citealt{Mouschovias_1976}), gravity begins to overcome the magnetic pressure in the cross-field direction, where the B-field provides the main support. Ambipolar diffusion within the dense condensations further weakens magnetic support, allowing these regions to become self-gravitating. As the dense clump undergoes gravitational collapse, it locally distorts and bends the B-field lines, producing a pinched, hourglass-like morphology. These condensations then continue to collapse to form stars, with their evolution still partially regulated by the surrounding B-field. This scenario is supported by the mass-to-flux ratio analysis of the G351 cloud (see Figure~\ref{fg5}), which shows that the cloud is magnetically transcritical on large scales, but becomes magnetically critical ($\lambda > 1$) near c1. Since the clump is massive enough to overcome the magnetic support, it distorts the initially perpendicular B-field lines, resulting in an hourglass-shaped morphology. Furthermore, the B-fields provide additional support that suppresses the formation of low-mass stars, allowing material to accumulate and eventually favoring the formation of massive stars. This B-field morphology observed in the G351 cloud is also supported by results from other MHD simulations as well. 
\citet{Gomez_2018} demonstrated that B-field lines initially oriented perpendicular to a filament can be bent by gas flows driven by self-gravity, giving rise to a characteristic `U'-shaped structure along the filament. In the case of c1, which is accreting material along the filament from both sides, these converging gas flows bend the field lines inward from both directions, resulting in an hourglass-shaped B-field morphology. 
A longitudinal velocity gradient along a filament is an observational signature of such mass accretion \citep{Mallick_2023}. However, no such gradient is detected in the C$^{18}$O(2--1) data for the G351 filament, likely due to the filament’s small projection along the line of sight.

Overall, the B-field plays a dominant role in regulating the dynamics of the G351 cloud, while gravitational forces locally distort the field lines, producing the observed hourglass-shaped magnetic morphology.

\subsection{Hourglass-shaped B-field due to converging sub-filaments}
\label{sec:disc2}

\begin{figure*}
	\centering
	\includegraphics[width=15cm]{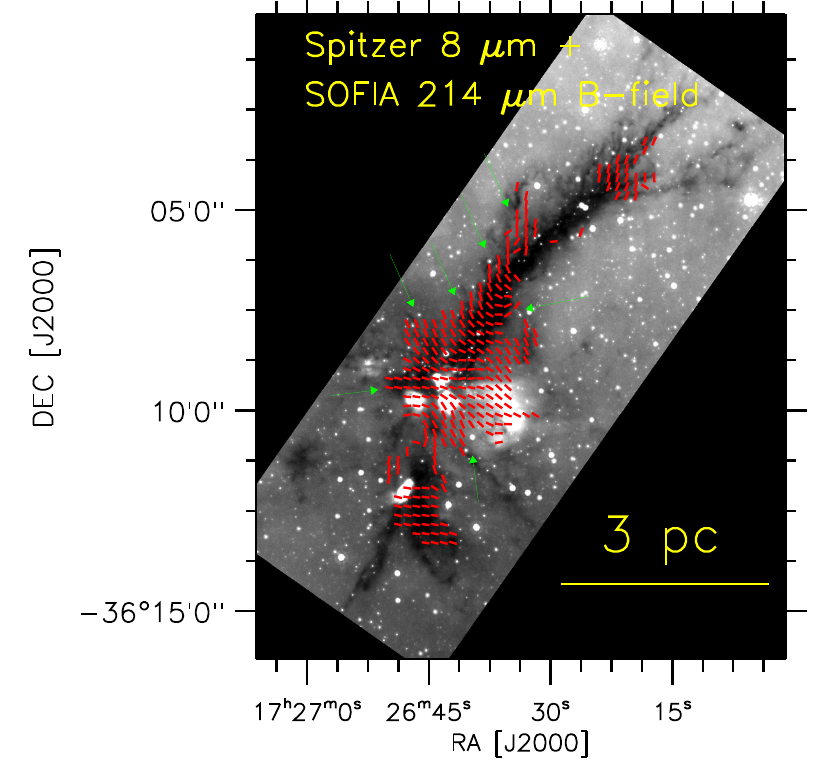}
	\caption{\textit{Spitzer} 8.0 $\mu$m image overlaid with the segments derived from the SOFIA/HAWC+ 214 $\mu$m polarization data. The green arrows highlight the sub-filaments aligned with the local B-field orientations. The scale bar corresponds to 3 pc at a distance of 2.0 kpc.}
	\label{fg6}
\end{figure*}

Considering that the magnetic and gravitational energies in the G351 cloud are comparable, the observed hourglass B-field morphology may also be influenced by the converging sub-filaments. \citet{Leurini_2019} reported multiple sub-filaments feeding into the central filamentary structure, consistent with accretion from the surrounding medium. 
In general, the B-field lines tends to align parallel to low-density filaments (or striations), while it becomes perpendicular to the higher-density filaments. This transition in field orientation typically occurs at column densities of about $N(\mathrm{H_2})$ $\sim$5 $\times$ $10^{21}$ $\mathrm{cm^{-2}}$. This behavior has been confirmed observationally and is also reproduced in MHD simulations \citep{Soler_2013, Planck_2015, Planck_2016, Soler_2017}. Figure~\ref{fg6} shows the SOFIA/HAWC+ 214 $\mu$m B-field vectors overlaid on the \textit{Spitzer} 8.0 $\mu$m image. A closer inspection shows that the B-field lines are aligned with the sub-filaments surrounding the central filament. While, the field lines are perpendicular to high density regions of filament. A similar signature is also observed in the Musca filament where the B-field lines are perpendicular to high-density filament, while parallel to the surrounding low-density striations \citep{Cox_2016}. In this scenario, the alignment of the sub-filaments with the B-field suggests that material is being accreted preferentially along the field lines, indicating B-field channeled mass inflow toward the central filament. Because the sub-filaments in the G351 cloud are not perpendicular to the main filament, the B-field remains aligned with the low-density striations but turns perpendicular within the higher-density regions (see green arrows in Fig:~\ref{fg6}). This transition in field orientation naturally give rise to an hourglass-shaped field morphology. A similar pattern is seen near the massive star-forming region around clump c1, where the `U'-shaped structure on either side of the clump shows a resemblance to an hourglass morphology. In this case, however, the apparent hourglass shape likely arises from the convergence of the sub-filaments rather than from a classical, gravitationally induced pinching of the B-field.
\citep{Leurini_2019} investigated the G351 filament using \textit{Herschel} multi-wavelength data and identified several sub-filaments converging toward the central filament and clump c1. Such a configuration is characteristic of a hub-filament system (HFS), in which multiple low aspect-ratio filaments feed material into a dense central hub where massive stars form \citep{Myers_2009}. The B-field lines in this study are aligned with the sub-filaments surrounding the clump c1, which is active masssive star-forming region in the G351 filament. A similar behaviour is also reported in Serpens South by \citet{Pillai_2020}, where the B-field lines are observed to follow the inflowing filaments and guide material toward the main hub. 

Given the limitations of the current datasets, it is not possible to determine which physical process predominantly drives the hourglass-shaped B-field geometry toward c1. It remains plausible that both gravitational contraction and magnetically guided inflow along sub-filaments act together to produce the observed morphology. Hence, high-resolution polarimetric and spectral-line observations toward c1 are required to distinguish between these scenarios.


\subsection{Comparison with other filamentary clouds and important caveats}

The B-field morphology observed in the G351 filamentary cloud shows a striking resemblance to the B-field structure in the Orion A/OMC-1 filament \citep[e.g.,][]{Pattle_2017,Chuss_2019,Hwang_2021}, as revealed by the JCMT and SOFIA observations. Previous studies of the OMC-1 filament have shown that the B-field lines are predominantly perpendicular to the filament, whereas toward the dense regions, the field exhibits an hourglass-like morphology. The mass-to-flux ratio and energy analyses indicate that the magnetic energy dominates the overall energy budget, followed by turbulence and gravity. However, within the dense regions, the relative importance of gravity increases \citep{Pattle_2017, Hwang_2021}.
The overall characteristics of B-field in OMC-1 are remarkably similar to those observed in the G351 cloud in this study. A similar B-field configuration has been reported in the massive DR21 filament using JCMT/POL-2 and SCUBA-2 observations \citep{Ching_2022}. In this filament, the B-field lines are predominantly perpendicular to the main pc-scale filament, while toward the massive dense core in the southern region, the field exhibits an hourglass-like morphology. %
Although similar B-field configurations have been observed in other filaments such as OMC-1 and DR21, reports of pinched, hourglass-shaped B-fields in IRDCs are limited, highlighting the critical role of B-fields in regulating the physical processes in the early stages of star formation.

%

 
The $B_{\rm POS}$  and $\lambda$ values estimated using the DCF method carry significant uncertainties, potentially up to a factor of two or more. The main contribution comes from the value of correction factor $Q$ in equation~\ref{eq1}. Numerous 3D MHD simulations have shown that this correction factor can vary from 0.2--1 depending on the turbulence regime, cloud geometry, and projection angle \citep{Liu_2021,Chen_2021,Myers_2024}. In particular, studies indicate that in strongly magnetized systems, the basic assumptions of the DCF method—namely small angular perturbations and approximate equipartition between turbulent kinetic and magnetic energy are better satisfied, and the correction factor approaches unity. In contrast, in weakly magnetized or self-gravitating systems, departures from equipartition, enhanced line-of-sight averaging, and unresolved field curvature lead to systematically smaller correction factors, typically $Q$ $\sim$ 0.3--0.4. Additionally, assumptions such as the cylindrical geometry of the filament cannot be precisely quantified and may contribute further to the uncertainties in the estimated $B_{\rm POS}$ values. 
The kinetic or turbulent energy can arise from two main sources: the cloud's intrinsic turbulence and the turbulence injected by stellar feedback. Consequently, it is difficult to disentangle the contributions of these two factors. Additionally, projection effects, including the unknown filament inclination and line-of-sight averaging, can bias the POS B-field strength measurements. In cases where the B-field strength is overestimated, the filament would transition from a magnetically transcritical state to a marginally supercritical regime. We therefore emphasize that the DCF-based estimates presented here should be interpreted as order-of-magnitude constraints, and that G351 lies close to the critical mass-to-flux threshold within the combined observational and methodological uncertainties.

\section{Summary and Conclusion}
\label{sec:summary}

We present a multi-wavelength study of the G351 IRDC, including the SOFIA/HAWC+ polarimetric observations at 214 $\mu$m. The key findings of this study are summarized as follows:

\begin{enumerate}
	\item A total of 53 Class I YSOs are identified toward the G351 cloud using the [4.5]$-$[5.8] vs [3.6]$-$[4.5] CCD. Most of these YSOs are located along the filamentary cloud, confirming active star formation within the region.
	\\
	\item The POS B-field morphology inferred from the SOFIA/HAWC+ 214 $\mu$m polarization data reveal that the B-field orientations are predominantly perpendicular to the major axis of the G351 filament.
	\\
	\item The B-field strength toward the G351 filament, estimated using the DCF method, is $\sim$147$\pm$ 60 $\mu$G. The mass-to-flux ratio analyses indicate that the filament is in a magnetically transcritical state ($\lambda$$\sim$1), where the gravitational and magnetic energies are comparable.
	\\
	\item The energy budget analysis reveals that the magnetic, kinetic, and gravitational energies contribute comparably in the G351 cloud,
	\\
	\item The POS B-field orientations toward the clump c1 reveals an hourglass-shaped morphology, with the B-field strengths values reaching upto $\sim$0.8 mG.
	\\
	\item The hourglass-shaped B-field toward c1 may arise from magnetically regulated gravitational collapse, from magnetically aligned converging sub-filaments, or from a combination of both processes.
\end{enumerate}

Overall, our results show that the B-fields are dynamically important in regulating the massive star formation activity in the G351 filamentary cloud. 

\section*{Acknowledgments}
The research work at Physical Research Laboratory is funded by the Department of Space, Government of India. IIZ is supported by the IAP RAS project FFUF-2024-0028. This publication is based on data acquired with the Atacama Pathfinder Experiment (APEX) under the programme ID 0104.F-9301(A). 
APEX is a collaboration between the Max-Planck-Institut fur Radioastronomie, the European Southern Observatory, and the Onsala Space Observatory. PS was partially supported by a Grant-in-Aid for Scientific Research (KAKENHI Number JP23H01221) of JSPS. RKY gratefully acknowledges support from the Fundamental Fund of Thailand Science Research and Innovation (TSRI) through the National Astronomical Research Institute of Thailand (Public Organization) (FFB680072/0269). This research has made use of the NASA/IPAC Infrared Science Archive, which is funded by the National Aeronautics and Space Administration and operated by the California Institute of Technology. The MeerKAT telescope is operated by the South African Radio Astronomy Observatory, which is a facility of the National Research Foundation, an agency of the Department of Science and Innovation. 
%
%
\subsection*{Data availability}
The {\it Herschel}, {\it Spitzer} and SOFIA data underlying this article are available from the publicly accessible NASA/IPAC infrared science archive\footnote{https://irsa.ipac.caltech.edu/frontpage/}.
The unWISE data underlying this article are available from the publicly accessible website\footnote{https://unwise.me/}.
The {\it Herschel} column density and temperature maps underlying this article are available from the publicly accessible website\footnote{http://www.astro.cardiff.ac.uk/research/ViaLactea/}. 
The SARAO MeerKAT 1.3 GHz continuum data underlying this article are available from the publicly accessible server\footnote{https://archive-gw-1.kat.ac.za/public/repository/10.48479/3wfd-e270/index.html/}. 
The ATLASGAL 870 $\mu$m dust continuum clumps underlying this article are available from the publicly accessible VizieR archive for catalogues\footnote{https://cdsarc.cds.unistra.fr/viz-bin/cat/}.
%


\bibliographystyle{mnras}
\bibliography{reference} 

@Article{	  Chambers_2009,
  author	= {{Chambers}, E.~T. and {Jackson}, J.~M. and {Rathborne},
		  J.~M. and {Simon}, R.},
  title		= "{Star Formation Activity of Cores within Infrared Dark
		  Clouds}",
  journal	= {\apjs},
  year		= 2009,
  month		= apr,
  volume	= {181},
  number	= {2},
  pages		= {360-390},
  doi		= {10.1088/0067-0049/181/2/360},
  adsurl	= {https://ui.adsabs.harvard.edu/abs/2009ApJS..181..360C}
}

@Article{	  Sanhueza_2012,
  author	= {{Sanhueza}, Patricio and {Jackson}, James M. and {Foster},
		  Jonathan B. and {Garay}, Guido and {Silva}, Andrea and
		  {Finn}, Susanna C.},
  title		= "{Chemistry in Infrared Dark Cloud Clumps: A Molecular Line
		  Survey at 3 mm}",
  journal	= {\apj},
  year		= 2012,
  month		= sep,
  volume	= {756},
  number	= {1},
  eid		= {60},
  pages		= {60},
  doi		= {10.1088/0004-637X/756/1/60},
  archiveprefix	= {arXiv},
  eprint	= {1206.6500},
  primaryclass	= {astro-ph.GA},
  adsurl	= {https://ui.adsabs.harvard.edu/abs/2012ApJ...756...60S}
}

@Article{	  Garrido_2025,
  author	= {{Sandoval-Garrido}, N.~A. and {Stutz}, A.~M. and
		  {{\'A}lvarez-Guti{\'e}rrez}, R.~H. and {Galv{\'a}n-Madrid},
		  R. and {Motte}, F. and {Ginsburg}, A. and {Cunningham}, N.
		  and {Reyes-Reyes}, S. and {Redaelli}, E. and {Bonfand}, M.
		  and {Salinas}, J. and {Koley}, A. and {Bernal-Mesina}, G.
		  and {Braine}, J. and {Bronfman}, L. and {Busquet}, G. and
		  {Csengeri}, T. and {Di Francesco}, J. and
		  {Fern{\'a}ndez-L{\'o}pez}, M. and {Garcia}, P. and
		  {Gusdorf}, A. and {Liu}, H.-L. and {Sanhueza}, P.},
  title		= "{ALMA-IMF: XVIII. The assembly of a star cluster: Dense
		  N$_{2}$H$^{+}$ (1{\textendash}0) kinematics in the massive
		  G351.77 protocluster}",
  journal	= {\aap},
  year		= 2025,
  month		= apr,
  volume	= {696},
  eid		= {A202},
  pages		= {A202},
  doi		= {10.1051/0004-6361/202452589},
  archiveprefix	= {arXiv},
  eprint	= {2410.09843},
  primaryclass	= {astro-ph.GA},
  adsurl	= {https://ui.adsabs.harvard.edu/abs/2025A&A...696A.202S}
}

@Article{	  Ishihara_2024,
  author	= {{Ishihara}, Kousuke and {Sanhueza}, Patricio and
		  {Nakamura}, Fumitaka and {Saito}, Masao and {Chen}, Huei-Ru
		  Vivien and {Li}, Shanghuo and {Olguin}, Fernando and
		  {Taniguchi}, Kotomi and {Morii}, Kaho and {Lu}, Xing and
		  {Luo}, Qiu-yi and {Sakai}, Takeshi and {Zhang}, Qizhou},
  title		= "{Digging into the Interior of Hot Cores with ALMA (DIHCA).
		  IV. Fragmentation in High-mass Star-forming Clumps}",
  journal	= {\apj},
  year		= 2024,
  month		= oct,
  volume	= {974},
  number	= {1},
  eid		= {95},
  pages		= {95},
  doi		= {10.3847/1538-4357/ad630f},
  archiveprefix	= {arXiv},
  eprint	= {2407.06845},
  primaryclass	= {astro-ph.GA},
  adsurl	= {https://ui.adsabs.harvard.edu/abs/2024ApJ...974...95I}
}

@Article{	  Ginsburg_2023,
  author	= {{Ginsburg}, Adam and {McGuire}, Brett A. and {Sanhueza},
		  Patricio and {Olguin}, Fernando and {Maud}, Luke T. and
		  {Tanaka}, Kei E.~I. and {Zhang}, Yichen and {Beuther},
		  Henrik and {Indriolo}, Nick},
  title		= "{Salt-bearing Disk Candidates around High-mass Young
		  Stellar Objects}",
  journal	= {\apj},
  year		= 2023,
  month		= jan,
  volume	= {942},
  number	= {2},
  eid		= {66},
  pages		= {66},
  doi		= {10.3847/1538-4357/ac9f4a},
  archiveprefix	= {arXiv},
  eprint	= {2211.02502},
  primaryclass	= {astro-ph.GA},
  adsurl	= {https://ui.adsabs.harvard.edu/abs/2023ApJ...942...66G}
}

@Article{	  Magnum_2015,
  author	= {{Mangum}, Jeffrey G. and {Shirley}, Yancy L.},
  title		= "{How to Calculate Molecular Column Density}",
  journal	= {\pasp},
  year		= 2015,
  month		= mar,
  volume	= {127},
  number	= {949},
  pages		= {266},
  doi		= {10.1086/680323},
  archiveprefix	= {arXiv},
  eprint	= {1501.01703},
  primaryclass	= {astro-ph.IM},
  adsurl	= {https://ui.adsabs.harvard.edu/abs/2015PASP..127..266M}
}

@Article{	  Yuan_2016,
  author	= {{Yuan}, Jinghua and {Wu}, Yuefang and {Liu}, Tie and
		  {Zhang}, Tianwei and {Zeng Li}, Jin and {Liu}, Hong-Li and
		  {Meng}, Fanyi and {Chen}, Ping and {Hu}, Runjie and {Wang},
		  Ke},
  title		= "{Dense Gas in Molecular Cores Associated with Planck
		  Galactic Cold Clumps}",
  journal	= {\apj},
  year		= 2016,
  month		= mar,
  volume	= {820},
  number	= {1},
  eid		= {37},
  pages		= {37},
  doi		= {10.3847/0004-637X/820/1/37},
  archiveprefix	= {arXiv},
  eprint	= {1601.04783},
  primaryclass	= {astro-ph.GA},
  adsurl	= {https://ui.adsabs.harvard.edu/abs/2016ApJ...820...37Y}
}

@Article{	  Pickett_1998,
  author	= {{Pickett}, H.~M. and {Poynter}, R.~L. and {Cohen}, E.~A.
		  and {Delitsky}, M.~L. and {Pearson}, J.~C. and
		  {M{\"u}ller}, H.~S.~P.},
  title		= "{Submillimeter, millimeter and microwave spectral line
		  catalog.}",
  journal	= {\jqsrt},
  year		= 1998,
  month		= nov,
  volume	= {60},
  number	= {5},
  pages		= {883-890},
  doi		= {10.1016/S0022-4073(98)00091-0},
  adsurl	= {https://ui.adsabs.harvard.edu/abs/1998JQSRT..60..883P}
}

@Article{	  Chandrasekhar_1953,
  author	= {{Chandrasekhar}, S. and {Fermi}, E.},
  title		= "{Magnetic Fields in Spiral Arms.}",
  journal	= {\apj},
  year		= 1953,
  month		= jul,
  volume	= {118},
  pages		= {113},
  doi		= {10.1086/145731},
  adsurl	= {https://ui.adsabs.harvard.edu/abs/1953ApJ...118..113C}
}

@Article{	  Crutcher_2012,
  author	= {{Crutcher}, Richard M.},
  title		= "{Magnetic Fields in Molecular Clouds}",
  journal	= {\araa},
  year		= 2012,
  month		= sep,
  volume	= {50},
  pages		= {29-63},
  doi		= {10.1146/annurev-astro-081811-125514},
  adsurl	= {https://ui.adsabs.harvard.edu/abs/2012ARA&A..50...29C}
}

@Article{	  Zucker_2018,
  author	= {{Zucker}, Catherine and {Chen}, Hope How-Huan},
  title		= "{RadFil: A Python Package for Building and Fitting Radial
		  Profiles for Interstellar Filaments}",
  journal	= {\apj},
  year		= 2018,
  month		= sep,
  volume	= {864},
  number	= {2},
  eid		= {152},
  pages		= {152},
  doi		= {10.3847/1538-4357/aad3b5},
  archiveprefix	= {arXiv},
  eprint	= {1807.06567},
  primaryclass	= {astro-ph.GA},
  adsurl	= {https://ui.adsabs.harvard.edu/abs/2018ApJ...864..152Z}
}

@InProceedings{Andre_2014,
  author	= {{Andr{\'e}}, P. and {Di Francesco}, J. and
		  {Ward-Thompson}, D. and {Inutsuka}, S. -I. and {Pudritz},
		  R.~E. and {Pineda}, J.~E.},
  title		= "{From Filamentary Networks to Dense Cores in Molecular
		  Clouds: Toward a New Paradigm for Star Formation}",
  booktitle	= {Protostars and Planets VI},
  year		= 2014,
  editor	= {{Beuther}, Henrik and {Klessen}, Ralf S. and {Dullemond},
		  Cornelis P. and {Henning}, Thomas},
  month		= jan,
  pages		= {27-51},
  doi		= {10.2458/azu_uapress_9780816531240-ch002},
  archiveprefix	= {arXiv},
  eprint	= {1312.6232},
  primaryclass	= {astro-ph.GA},
  adsurl	= {https://ui.adsabs.harvard.edu/abs/2014prpl.conf...27A}
}

@Article{	  Hildebrand_2009,
  author	= {{Hildebrand}, Roger H. and {Kirby}, Larry and {Dotson},
		  Jessie L. and {Houde}, Martin and {Vaillancourt}, John E.},
  title		= "{Dispersion of Magnetic Fields in Molecular Clouds. I}",
  journal	= {\apj},
  year		= 2009,
  month		= may,
  volume	= {696},
  number	= {1},
  pages		= {567-573},
  doi		= {10.1088/0004-637X/696/1/567},
  archiveprefix	= {arXiv},
  eprint	= {0811.0813},
  primaryclass	= {astro-ph},
  adsurl	= {https://ui.adsabs.harvard.edu/abs/2009ApJ...696..567H}
}

@Article{	  Houde_2009,
  author	= {{Houde}, Martin and {Vaillancourt}, John E. and
		  {Hildebrand}, Roger H. and {Chitsazzadeh}, Shadi and
		  {Kirby}, Larry},
  title		= "{Dispersion of Magnetic Fields in Molecular Clouds. II.}",
  journal	= {\apj},
  year		= 2009,
  month		= dec,
  volume	= {706},
  number	= {2},
  pages		= {1504-1516},
  doi		= {10.1088/0004-637X/706/2/1504},
  archiveprefix	= {arXiv},
  eprint	= {0909.5227},
  primaryclass	= {astro-ph.GA},
  adsurl	= {https://ui.adsabs.harvard.edu/abs/2009ApJ...706.1504H}
}

@Article{	  Hwang_2021,
  author	= {{Hwang}, Jihye and {Kim}, Jongsoo and {Pattle}, Kate and
		  {Kwon}, Woojin and {Sadavoy}, Sarah and {Koch}, Patrick M.
		  and {Hull}, Charles L.~H. and {Johnstone}, Doug and
		  {Furuya}, Ray S. and {Won Lee}, Chang and {Arzoumanian},
		  Doris and {Tahani}, Mehrnoosh and {Eswaraiah}, Chakali and
		  {Liu}, Tie and {Kirchschlager}, Florian and {Kim}, Kee-Tae
		  and {Tamura}, Motohide and {Kwon}, Jungmi and {Lyo}, A.
		  -Ran and {Soam}, Archana and {Kang}, Ji-hyun and {Bourke},
		  Tyler L. and {Matsumura}, Masafumi and {Mairs}, Steve and
		  {Kim}, Gwanjeong and {Park}, Geumsook and {Nakamura},
		  Fumitaka and {Onaka}, Takashi and {Tang}, Xindi and {Liu},
		  Hong-Li and {Ward-Thompson}, Derek and {Li}, Di and
		  {Hoang}, Thiem and {Hasegawa}, Tetsuo and {Qiu}, Keping and
		  {Lai}, Shih-Ping and {Bastien}, Pierre},
  title		= "{The JCMT BISTRO Survey: The Distribution of Magnetic
		  Field Strengths toward the OMC-1 Region}",
  journal	= {\apj},
  year		= 2021,
  month		= jun,
  volume	= {913},
  number	= {2},
  eid		= {85},
  pages		= {85},
  doi		= {10.3847/1538-4357/abf3c4},
  archiveprefix	= {arXiv},
  eprint	= {2103.16144},
  primaryclass	= {astro-ph.GA},
  adsurl	= {https://ui.adsabs.harvard.edu/abs/2021ApJ...913...85H}
}

@Article{	  Guerra_2021,
  author	= {{Guerra}, Jordan A. and {Chuss}, David T. and {Dowell}, C.
		  Darren and {Houde}, Martin and {Michail}, Joseph M. and
		  {Siah}, Javad and {Wollack}, Edward J.},
  title		= "{Maps of Magnetic Field Strength in the OMC-1 Using HAWC+
		  FIR Polarimetric Data}",
  journal	= {\apj},
  year		= 2021,
  month		= feb,
  volume	= {908},
  number	= {1},
  eid		= {98},
  pages		= {98},
  doi		= {10.3847/1538-4357/abd6f0},
  archiveprefix	= {arXiv},
  eprint	= {2007.04923},
  primaryclass	= {astro-ph.GA},
  adsurl	= {https://ui.adsabs.harvard.edu/abs/2021ApJ...908...98G}
}

@Article{	  Ngan_2024,
  author	= {{L{\^e}}, Ng{\^a}n and {Tram}, Le Ngoc and {Karska}, Agata
		  and {Hoang}, Thiem and {Diep}, Pham Ngoc and {Hanasz},
		  Micha{\l} and {Ngoc}, Nguyen Bich and {Phuong}, Nguyen Thi
		  and {Menten}, Karl M. and {Wyrowski}, Friedrich and
		  {Nguyen}, Dieu D. and {Hoang}, Thuong Duc and {Khang},
		  Nguyen Minh},
  title		= "{Mapping and characterizing magnetic fields in the Rho
		  Ophiuchus-A molecular cloud with SOFIA/HAWC+}",
  journal	= {\aap},
  year		= 2024,
  month		= oct,
  volume	= {690},
  eid		= {A191},
  pages		= {A191},
  doi		= {10.1051/0004-6361/202348008},
  archiveprefix	= {arXiv},
  eprint	= {2408.17122},
  primaryclass	= {astro-ph.GA},
  adsurl	= {https://ui.adsabs.harvard.edu/abs/2024A&A...690A.191L}
}

@Article{	  Fuller_1992,
  author	= {{Fuller}, G.~A. and {Myers}, P.~C.},
  title		= "{Dense Cores in Dark Clouds. VII. Line Width--Size
		  Relations}",
  journal	= {\apj},
  year		= 1992,
  month		= jan,
  volume	= {384},
  pages		= {523},
  doi		= {10.1086/170894},
  adsurl	= {https://ui.adsabs.harvard.edu/abs/1992ApJ...384..523F}
}

@Article{	  Kauffmann_2013,
  author	= {{Kauffmann}, Jens and {Pillai}, Thushara and {Goldsmith},
		  Paul F.},
  title		= "{Low Virial Parameters in Molecular Clouds: Implications
		  for High-mass Star Formation and Magnetic Fields}",
  journal	= {\apj},
  year		= 2013,
  month		= dec,
  volume	= {779},
  number	= {2},
  eid		= {185},
  pages		= {185},
  doi		= {10.1088/0004-637X/779/2/185},
  archiveprefix	= {arXiv},
  eprint	= {1308.5679},
  primaryclass	= {astro-ph.GA},
  adsurl	= {https://ui.adsabs.harvard.edu/abs/2013ApJ...779..185K}
}

@Article{	  Hartmann_2005,
  author	= {{Hartmann}, Lee and {Megeath}, S.~T. and {Allen}, Lori and
		  {Luhman}, Kevin and {Calvet}, Nuria and {D'Alessio}, Paola
		  and {Franco-Hernandez}, Ramiro and {Fazio}, Giovanni},
  title		= "{IRAC Observations of Taurus Pre-Main-Sequence Stars}",
  journal	= {\apj},
  year		= 2005,
  month		= aug,
  volume	= {629},
  number	= {2},
  pages		= {881-896},
  doi		= {10.1086/431472},
  archiveprefix	= {arXiv},
  eprint	= {astro-ph/0505323},
  primaryclass	= {astro-ph},
  adsurl	= {https://ui.adsabs.harvard.edu/abs/2005ApJ...629..881H}
}

@Article{	  Getman_2007,
  author	= {{Getman}, Konstantin V. and {Feigelson}, Eric D. and
		  {Garmire}, Gordon and {Broos}, Patrick and {Wang},
		  Junfeng},
  title		= "{X-Ray Study of Triggered Star Formation and Protostars in
		  IC 1396N}",
  journal	= {\apj},
  year		= 2007,
  month		= jan,
  volume	= {654},
  number	= {1},
  pages		= {316-337},
  doi		= {10.1086/509112},
  archiveprefix	= {arXiv},
  eprint	= {astro-ph/0607006},
  primaryclass	= {astro-ph},
  adsurl	= {https://ui.adsabs.harvard.edu/abs/2007ApJ...654..316G}
}

@Article{	  Dewangan_2015,
  author	= {{Dewangan}, L.~K. and {Luna}, A. and {Ojha}, D.~K. and
		  {Anandarao}, B.~G. and {Mallick}, K.~K. and {Mayya},
		  Y.~D.},
  title		= "{The Physical Environment of the Massive Star-forming
		  Region W42}",
  journal	= {\apj},
  year		= 2015,
  month		= oct,
  volume	= {811},
  number	= {2},
  eid		= {79},
  pages		= {79},
  doi		= {10.1088/0004-637X/811/2/79},
  archiveprefix	= {arXiv},
  eprint	= {1508.04425},
  primaryclass	= {astro-ph.GA},
  adsurl	= {https://ui.adsabs.harvard.edu/abs/2015ApJ...811...79D}
}

@Article{	  Crutcher_2004,
  author	= {{Crutcher}, Richard M.},
  title		= "{What Drives Star Formation?}",
  journal	= {\apss},
  year		= 2004,
  month		= aug,
  volume	= {292},
  number	= {1},
  pages		= {225-237},
  doi		= {10.1023/B:ASTR.0000045021.42255.95},
  adsurl	= {https://ui.adsabs.harvard.edu/abs/2004Ap&SS.292..225C}
}

@Article{	  Fiege_2000,
  author	= {{Fiege}, Jason D. and {Pudritz}, Ralph E.},
  title		= "{Polarized Submillimeter Emission from Filamentary
		  Molecular Clouds}",
  journal	= {\apj},
  year		= 2000,
  month		= dec,
  volume	= {544},
  number	= {2},
  pages		= {830-837},
  doi		= {10.1086/317228},
  archiveprefix	= {arXiv},
  eprint	= {astro-ph/0005363},
  primaryclass	= {astro-ph},
  adsurl	= {https://ui.adsabs.harvard.edu/abs/2000ApJ...544..830F}
}

@Article{	  Sanhueza_2025,
  author	= {{Sanhueza}, Patricio and {Liu}, Junhao and {Morii}, Kaho
		  and {Girart}, Josep Miquel and {Zhang}, Qizhou and
		  {Stephens}, Ian W. and {Jackson}, James M. and
		  {Cort{\'e}s}, Paulo C. and {Koch}, Patrick M. and
		  {Cyganowski}, Claudia J. and {Saha}, Piyali and {Beuther},
		  Henrik and {Zhang}, Suinan and {Beltr{\'a}n}, Maria T. and
		  {Cheng}, Yu and {Olguin}, Fernando A. and {Lu}, Xing and
		  {Choudhury}, Spandan and {Pattle}, Kate and
		  {Fern{\'a}ndez-L{\'o}pez}, Manuel and {Hwang}, Jihye and
		  {Kang}, Ji-hyun and {Karoly}, Janik and {Ginsburg}, Adam
		  and {Lyo}, A. -Ran and {Taniguchi}, Kotomi and {Jiao},
		  Wenyu and {Eswaraiah}, Chakali and {Luo}, Qiu-yi and
		  {Wang}, Jia-Wei and {Commer{\c{c}}on}, Beno{\^\i}t and
		  {Li}, Shanghuo and {Xu}, Fengwei and {Chen}, Huei-Ru Vivien
		  and {Zapata}, Luis A. and {Chung}, Eun Jung and {Nakamura},
		  Fumitaka and {Panigrahy}, Sandhyarani and {Sakai}, Takeshi},
  title		= "{Magnetic Fields in Massive Star-forming Regions (MagMaR).
		  V. The Magnetic Field at the Onset of High-mass Star
		  Formation}",
  journal	= {\apj},
  year		= 2025,
  month		= feb,
  volume	= {980},
  number	= {1},
  eid		= {87},
  pages		= {87},
  doi		= {10.3847/1538-4357/ad9d40},
  archiveprefix	= {arXiv},
  eprint	= {2412.08790},
  primaryclass	= {astro-ph.GA},
  adsurl	= {https://ui.adsabs.harvard.edu/abs/2025ApJ...980...87S}
}

@Article{	  Arzoumanian_2011,
  author	= {{Arzoumanian}, D. and {Andr{\'e}}, Ph. and {Didelon}, P.
		  and {K{\"o}nyves}, V. and {Schneider}, N. and
		  {Men'shchikov}, A. and {Sousbie}, T. and {Zavagno}, A. and
		  {Bontemps}, S. and {di Francesco}, J. and {Griffin}, M. and
		  {Hennemann}, M. and {Hill}, T. and {Kirk}, J. and {Martin},
		  P. and {Minier}, V. and {Molinari}, S. and {Motte}, F. and
		  {Peretto}, N. and {Pezzuto}, S. and {Spinoglio}, L. and
		  {Ward-Thompson}, D. and {White}, G. and {Wilson}, C.~D.},
  title		= "{Characterizing interstellar filaments with Herschel in IC
		  5146}",
  journal	= {\aap},
  year		= 2011,
  month		= may,
  volume	= {529},
  eid		= {L6},
  pages		= {L6},
  doi		= {10.1051/0004-6361/201116596},
  archiveprefix	= {arXiv},
  eprint	= {1103.0201},
  primaryclass	= {astro-ph.GA},
  adsurl	= {https://ui.adsabs.harvard.edu/abs/2011A&A...529L...6A}
}

@Article{	  Rathborne_2006,
  author	= {{Rathborne}, J.~M. and {Jackson}, J.~M. and {Simon}, R.},
  title		= "{Infrared Dark Clouds: Precursors to Star Clusters}",
  journal	= {\apj},
  year		= 2006,
  month		= apr,
  volume	= {641},
  number	= {1},
  pages		= {389-405},
  doi		= {10.1086/500423},
  archiveprefix	= {arXiv},
  eprint	= {astro-ph/0602246},
  primaryclass	= {astro-ph},
  adsurl	= {https://ui.adsabs.harvard.edu/abs/2006ApJ...641..389R}
}

@Article{	  Planck_2016,
  author	= {{Planck Collaboration} and {Ade}, P.~A.~R. and {Aghanim},
		  N. and {Alves}, M.~I.~R. and {Arnaud}, M. and
		  {Arzoumanian}, D. and {Ashdown}, M. and {Aumont}, J. and
		  {Baccigalupi}, C. and {Banday}, A.~J. and {Barreiro}, R.~B.
		  and {Bartolo}, N. and {Battaner}, E. and {Benabed}, K. and
		  {Beno{\^\i}t}, A. and {Benoit-L{\'e}vy}, A. and {Bernard},
		  J. -P. and {Bersanelli}, M. and {Bielewicz}, P. and {Bock},
		  J.~J. and {Bonavera}, L. and {Bond}, J.~R. and {Borrill},
		  J. and {Bouchet}, F.~R. and {Boulanger}, F. and {Bracco},
		  A. and {Burigana}, C. and {Calabrese}, E. and {Cardoso}, J.
		  -F. and {Catalano}, A. and {Chiang}, H.~C. and
		  {Christensen}, P.~R. and {Colombo}, L.~P.~L. and {Combet},
		  C. and {Couchot}, F. and {Crill}, B.~P. and {Curto}, A. and
		  {Cuttaia}, F. and {Danese}, L. and {Davies}, R.~D. and
		  {Davis}, R.~J. and {de Bernardis}, P. and {de Rosa}, A. and
		  {de Zotti}, G. and {Delabrouille}, J. and {Dickinson}, C.
		  and {Diego}, J.~M. and {Dole}, H. and {Donzelli}, S. and
		  {Dor{\'e}}, O. and {Douspis}, M. and {Ducout}, A. and
		  {Dupac}, X. and {Efstathiou}, G. and {Elsner}, F. and
		  {En{\ss}lin}, T.~A. and {Eriksen}, H.~K. and
		  {Falceta-Gon{\c{c}}alves}, D. and {Falgarone}, E. and
		  {Ferri{\`e}re}, K. and {Finelli}, F. and {Forni}, O. and
		  {Frailis}, M. and {Fraisse}, A.~A. and {Franceschi}, E. and
		  {Frejsel}, A. and {Galeotta}, S. and {Galli}, S. and
		  {Ganga}, K. and {Ghosh}, T. and {Giard}, M. and
		  {Gjerl{\o}w}, E. and {Gonz{\'a}lez-Nuevo}, J. and
		  {G{\'o}rski}, K.~M. and {Gregorio}, A. and {Gruppuso}, A.
		  and {Gudmundsson}, J.~E. and {Guillet}, V. and {Harrison},
		  D.~L. and {Helou}, G. and {Hennebelle}, P. and
		  {Henrot-Versill{\'e}}, S. and {Hern{\'a}ndez-Monteagudo},
		  C. and {Herranz}, D. and {Hildebrandt}, S.~R. and {Hivon},
		  E. and {Holmes}, W.~A. and {Hornstrup}, A. and
		  {Huffenberger}, K.~M. and {Hurier}, G. and {Jaffe}, A.~H.
		  and {Jaffe}, T.~R. and {Jones}, W.~C. and {Juvela}, M. and
		  {Keih{\"a}nen}, E. and {Keskitalo}, R. and {Kisner}, T.~S.
		  and {Knoche}, J. and {Kunz}, M. and {Kurki-Suonio}, H. and
		  {Lagache}, G. and {Lamarre}, J. -M. and {Lasenby}, A. and
		  {Lattanzi}, M. and {Lawrence}, C.~R. and {Leonardi}, R. and
		  {Levrier}, F. and {Liguori}, M. and {Lilje}, P.~B. and
		  {Linden-V{\o}rnle}, M. and {L{\'o}pez-Caniego}, M. and
		  {Lubin}, P.~M. and {Mac{\'\i}as-P{\'e}rez}, J.~F. and
		  {Maino}, D. and {Mandolesi}, N. and {Mangilli}, A. and
		  {Maris}, M. and {Martin}, P.~G. and
		  {Mart{\'\i}nez-Gonz{\'a}lez}, E. and {Masi}, S. and
		  {Matarrese}, S. and {Melchiorri}, A. and {Mendes}, L. and
		  {Mennella}, A. and {Migliaccio}, M. and
		  {Miville-Desch{\^e}nes}, M. -A. and {Moneti}, A. and
		  {Montier}, L. and {Morgante}, G. and {Mortlock}, D. and
		  {Munshi}, D. and {Murphy}, J.~A. and {Naselsky}, P. and
		  {Nati}, F. and {Netterfield}, C.~B. and {Noviello}, F. and
		  {Novikov}, D. and {Novikov}, I. and {Oppermann}, N. and
		  {Oxborrow}, C.~A. and {Pagano}, L. and {Pajot}, F. and
		  {Paladini}, R. and {Paoletti}, D. and {Pasian}, F. and
		  {Perotto}, L. and {Pettorino}, V. and {Piacentini}, F. and
		  {Piat}, M. and {Pierpaoli}, E. and {Pietrobon}, D. and
		  {Plaszczynski}, S. and {Pointecouteau}, E. and {Polenta},
		  G. and {Ponthieu}, N. and {Pratt}, G.~W. and {Prunet}, S.
		  and {Puget}, J. -L. and {Rachen}, J.~P. and {Reinecke}, M.
		  and {Remazeilles}, M. and {Renault}, C. and {Renzi}, A. and
		  {Ristorcelli}, I. and {Rocha}, G. and {Rossetti}, M. and
		  {Roudier}, G. and {Rubi{\~n}o-Mart{\'\i}n}, J.~A. and
		  {Rusholme}, B. and {Sandri}, M. and {Santos}, D. and
		  {Savelainen}, M. and {Savini}, G. and {Scott}, D. and
		  {Soler}, J.~D. and {Stolyarov}, V. and {Sudiwala}, R. and
		  {Sutton}, D. and {Suur-Uski}, A. -S. and {Sygnet}, J. -F.
		  and {Tauber}, J.~A. and {Terenzi}, L. and {Toffolatti}, L.
		  and {Tomasi}, M. and {Tristram}, M. and {Tucci}, M. and
		  {Umana}, G. and {Valenziano}, L. and {Valiviita}, J. and
		  {Van Tent}, B. and {Vielva}, P. and {Villa}, F. and {Wade},
		  L.~A. and {Wandelt}, B.~D. and {Wehus}, I.~K. and {Ysard},
		  N. and {Yvon}, D. and {Zonca}, A.},
  title		= "{Planck intermediate results. XXXV. Probing the role of
		  the magnetic field in the formation of structure in
		  molecular clouds}",
  journal	= {\aap},
  year		= 2016,
  month		= feb,
  volume	= {586},
  eid		= {A138},
  pages		= {A138},
  doi		= {10.1051/0004-6361/201525896},
  archiveprefix	= {arXiv},
  eprint	= {1502.04123},
  primaryclass	= {astro-ph.GA},
  adsurl	= {https://ui.adsabs.harvard.edu/abs/2016A&A...586A.138P}
}

@Article{	  Soler_2013,
  author	= {{Soler}, J.~D. and {Hennebelle}, P. and {Martin}, P.~G.
		  and {Miville-Desch{\^e}nes}, M.-A. and {Netterfield}, C.~B.
		  and {Fissel}, L.~M.},
  title		= "{An Imprint of Molecular Cloud Magnetization in the
		  Morphology of the Dust Polarized Emission}",
  journal	= {\apj},
  year		= 2013,
  month		= sep,
  volume	= {774},
  number	= {2},
  eid		= {128},
  pages		= {128},
  doi		= {10.1088/0004-637X/774/2/128},
  archiveprefix	= {arXiv},
  eprint	= {1303.1830},
  primaryclass	= {astro-ph.GA},
  adsurl	= {https://ui.adsabs.harvard.edu/abs/2013ApJ...774..128S}
}

@Article{	  Nakamura_2008,
  author	= {{Nakamura}, Fumitaka and {Li}, Zhi-Yun},
  title		= "{Magnetically Regulated Star Formation in Three
		  Dimensions: The Case of the Taurus Molecular Cloud
		  Complex}",
  journal	= {\apj},
  year		= 2008,
  month		= nov,
  volume	= {687},
  number	= {1},
  pages		= {354-375},
  doi		= {10.1086/591641},
  archiveprefix	= {arXiv},
  eprint	= {0804.4201},
  primaryclass	= {astro-ph},
  adsurl	= {https://ui.adsabs.harvard.edu/abs/2008ApJ...687..354N}
}

@Article{	  Soler_2017,
  author	= {{Soler}, J.~D. and {Hennebelle}, P.},
  title		= "{What are we learning from the relative orientation
		  between density structures and the magnetic field in
		  molecular clouds?}",
  journal	= {\aap},
  year		= 2017,
  month		= oct,
  volume	= {607},
  eid		= {A2},
  pages		= {A2},
  doi		= {10.1051/0004-6361/201731049},
  archiveprefix	= {arXiv},
  eprint	= {1705.00477},
  primaryclass	= {astro-ph.GA},
  adsurl	= {https://ui.adsabs.harvard.edu/abs/2017A&A...607A...2S}
}

@Article{	  Seifred_2015,
  author	= {{Seifried}, D. and {Walch}, S.},
  title		= "{The impact of turbulence and magnetic field orientation
		  on star-forming filaments}",
  journal	= {\mnras},
  year		= 2015,
  month		= sep,
  volume	= {452},
  number	= {3},
  pages		= {2410-2422},
  doi		= {10.1093/mnras/stv1458},
  archiveprefix	= {arXiv},
  eprint	= {1503.01659},
  primaryclass	= {astro-ph.GA},
  adsurl	= {https://ui.adsabs.harvard.edu/abs/2015MNRAS.452.2410S}
}

@Article{	  Chen_2015,
  author	= {{Chen}, Che-Yu and {Ostriker}, Eve C.},
  title		= "{Anisotropic Formation of Magnetized Cores in Turbulent
		  Clouds}",
  journal	= {\apj},
  year		= 2015,
  month		= sep,
  volume	= {810},
  number	= {2},
  eid		= {126},
  pages		= {126},
  doi		= {10.1088/0004-637X/810/2/126},
  archiveprefix	= {arXiv},
  eprint	= {1508.02710},
  primaryclass	= {astro-ph.GA},
  adsurl	= {https://ui.adsabs.harvard.edu/abs/2015ApJ...810..126C}
}

@Article{	  Maity_2024,
  author	= {{Maity}, A.~K. and {Inoue}, T. and {Fukui}, Y. and
		  {Dewangan}, L.~K. and {Sano}, H. and {Yamada}, R.~I. and
		  {Tachihara}, K. and {Bhadari}, N.~K. and {Jadhav}, O.~R.},
  title		= "{Cloud{\textendash}Cloud Collision: Formation of
		  Hub-filament Systems and Associated Gas Kinematics.
		  Mass-collecting Cone{\textemdash}A New Signature of
		  Cloud{\textendash}Cloud Collision}",
  journal	= {\apj},
  year		= 2024,
  month		= oct,
  volume	= {974},
  number	= {2},
  eid		= {229},
  pages		= {229},
  doi		= {10.3847/1538-4357/ad7098},
  archiveprefix	= {arXiv},
  eprint	= {2408.06826},
  primaryclass	= {astro-ph.GA},
  adsurl	= {https://ui.adsabs.harvard.edu/abs/2024ApJ...974..229M}
}

@Article{	  Ngoc_2023,
  author	= {{Ngoc}, Nguyen Bich and {Diep}, Pham Ngoc and {Hoang},
		  Thiem and {Tram}, Le Ngoc and {Giang}, Nguyen Chau and
		  {L{\^e}}, Ng{\^a}n and {Hoang}, Thuong D. and {Phuong},
		  Nguyen Thi and {Khang}, Nguyen Minh and {Nguyen}, Dieu D.
		  and {Truong}, Bao},
  title		= "{B-fields and Dust in Interstellar Filaments Using Dust
		  Polarization (BALLAD-POL). I. The Massive Filament
		  G11.11-0.12 Observed by SOFIA/HAWC+}",
  journal	= {\apj},
  year		= 2023,
  month		= aug,
  volume	= {953},
  number	= {1},
  eid		= {66},
  pages		= {66},
  doi		= {10.3847/1538-4357/acdb6e},
  archiveprefix	= {arXiv},
  eprint	= {2302.10543},
  primaryclass	= {astro-ph.GA},
  adsurl	= {https://ui.adsabs.harvard.edu/abs/2023ApJ...953...66N}
}

@Article{	  Chen_2023,
  author	= {{Chen}, Zhiwei and {Sefako}, Ramotholo and {Yang}, Yang
		  and {Jiang}, Zhibo and {Su}, Yang and {Zhang}, Shaobo and
		  {Zhou}, Xin},
  title		= "{The role of magnetic fields in the formation of the
		  filamentary infrared dark cloud G11.11-0.12}",
  journal	= {\mnras},
  year		= 2023,
  month		= oct,
  volume	= {525},
  number	= {1},
  pages		= {107-122},
  doi		= {10.1093/mnras/stad2259},
  archiveprefix	= {arXiv},
  eprint	= {2207.03695},
  primaryclass	= {astro-ph.GA},
  adsurl	= {https://ui.adsabs.harvard.edu/abs/2023MNRAS.525..107C}
}

@Article{	  Truong_2025,
  author	= {{Truong}, Bao and {Hoang}, Thiem and {Bich Ngoc}, Nguyen
		  and {Chau Giang}, Nguyen and {Tram}, Le Ngoc and {Le},
		  Ngan},
  title		= "{3D B-fieLds in the InterStellar medium and Star-forming
		  regions (3D-BLISS): I. Using Starlight Polarization in
		  Massive IRDC Filament G11.11-0.12}",
  journal	= {arXiv e-prints},
  year		= 2025,
  month		= oct,
  eid		= {arXiv:2510.06726},
  pages		= {arXiv:2510.06726},
  doi		= {10.48550/arXiv.2510.06726},
  archiveprefix	= {arXiv},
  eprint	= {2510.06726},
  primaryclass	= {astro-ph.GA},
  adsurl	= {https://ui.adsabs.harvard.edu/abs/2025arXiv251006726T}
}

@Article{	  Law_2024,
  author	= {{Law}, Chi-Yan and {Tan}, Jonathan C. and {Skalidis},
		  Raphael and {Morgan}, Larry and {Xu}, Duo and {de Oliveira
		  Alves}, Felipe and {Barnes}, Ashley T. and {Butterfield},
		  Natalie and {Caselli}, Paola and {Cosentino}, Giuliana and
		  {Fontani}, Francesco and {Henshaw}, Jonathan D. and
		  {Jimenez-Serra}, Izaskun and {Lim}, Wanggi},
  title		= "{Polarized Light from Massive Protoclusters (POLIMAP). I.
		  Dissecting the Role of Magnetic Fields in the Massive
		  Infrared Dark Cloud G28.37+0.07}",
  journal	= {\apj},
  year		= 2024,
  month		= jun,
  volume	= {967},
  number	= {2},
  eid		= {157},
  pages		= {157},
  doi		= {10.3847/1538-4357/ad39e0},
  archiveprefix	= {arXiv},
  eprint	= {2401.11560},
  primaryclass	= {astro-ph.GA},
  adsurl	= {https://ui.adsabs.harvard.edu/abs/2024ApJ...967..157L}
}

@Article{	  Hwang_2025,
  author	= {{Hwang}, Jihye and {Pattle}, Kate and {Lee}, Chang Won and
		  {Karoly}, Janik and {Kim}, Kee-Tae and {Kim}, Jongsoo and
		  {Liu}, Junhao and {Qiu}, Keping and {Lyo}, A.-Ran and
		  {Eden}, David and {Koch}, Patrick M. and {Arzoumanian},
		  Doris and {Sharma}, Ekta and {Poidevin}, Fr{\'e}d{\'e}rick
		  and {Johnstone}, Doug and {Coud{\'e}}, Simon and {Tahani},
		  Mehrnoosh and {Ward-Thompson}, Derek and {Soam}, Archana
		  and {Kang}, Ji-hyun and {Hoang}, Thiem and {Kwon}, Woojin
		  and {Ngoc}, Nguyen Bich and {Chung}, Eun Jung and {Bourke},
		  Tyler L. and {Onaka}, Takashi and {Kirchschlager}, Florian
		  and {Tamura}, Motohide and {Kwon}, Jungmi and {Tang}, Xindi
		  and {Chakali}, Eswaraiah and {Liu}, Tie and {Bastien},
		  Pierre and {Furuya}, Ray S. and {Lai}, Shih-Ping and {Lin},
		  Sheng-Jun and {Wang}, Jia-Wei and {Berry}, David},
  title		= "{The JCMT BISTRO-3 Survey: Variation of Magnetic Field
		  Orientations on Parsec and Subparsec Scales in the Massive
		  Star-forming Region G28.34+0.06}",
  journal	= {\apj},
  year		= 2025,
  month		= jun,
  volume	= {985},
  number	= {2},
  eid		= {222},
  pages		= {222},
  doi		= {10.3847/1538-4357/adce80},
  archiveprefix	= {arXiv},
  eprint	= {2505.14047},
  primaryclass	= {astro-ph.GA},
  adsurl	= {https://ui.adsabs.harvard.edu/abs/2025ApJ...985..222H}
}

@Article{	  Liu_2024,
  author	= {{Liu}, Junhao and {Zhang}, Qizhou and {Lin}, Yuxin and
		  {Qiu}, Keping and {Koch}, Patrick M. and {Liu}, Hauyu
		  Baobab and {Li}, Zhi-Yun and {Girart}, Josep Miquel and
		  {Pillai}, Thushara G.~S. and {Li}, Shanghuo and {Chen},
		  Huei-Ru Vivien and {Ching}, Tao-Chung and {Ho}, Paul T.~P.
		  and {Lai}, Shih-Ping and {Rao}, Ramprasad and {Tang},
		  Ya-Wen and {Wang}, Ke},
  title		= "{Dark Dragon Breaks Magnetic Chain: Dynamical
		  Substructures of IRDC G28.34 Form in Supported
		  Environments}",
  journal	= {\apj},
  year		= 2024,
  month		= may,
  volume	= {966},
  number	= {1},
  eid		= {120},
  pages		= {120},
  doi		= {10.3847/1538-4357/ad3105},
  archiveprefix	= {arXiv},
  eprint	= {2403.03437},
  primaryclass	= {astro-ph.GA},
  adsurl	= {https://ui.adsabs.harvard.edu/abs/2024ApJ...966..120L}
}

@Article{	  Pravash_2025,
  author	= {{Pravash}, Saikhom and {Soam}, Archana and {Diep}, Pham
		  Ngoc and {Hoang}, Thiem and {Ngoc}, Nguyen Bich and {Tram},
		  Le Ngoc},
  title		= "{B-fields and Dust in Interstellar Filaments Using Dust
		  Polarization (BALLAD-POL). III. Grain Alignment and
		  Disruption Mechanisms in G34.43+0.24 Using Polarization
		  Observations from JCMT/POL-2}",
  journal	= {\apj},
  year		= 2025,
  month		= mar,
  volume	= {981},
  number	= {2},
  eid		= {128},
  pages		= {128},
  doi		= {10.3847/1538-4357/adae06},
  archiveprefix	= {arXiv},
  eprint	= {2501.11634},
  primaryclass	= {astro-ph.GA},
  adsurl	= {https://ui.adsabs.harvard.edu/abs/2025ApJ...981..128P}
}

@Article{	  Soam_2019,
  author	= {{Soam}, Archana and {Liu}, Tie and {Andersson}, B.-G. and
		  {Lee}, Chang Won and {Liu}, Junhao and {Juvela}, Mika and
		  {Li}, Pak Shing and {Goldsmith}, Paul F. and {Zhang},
		  Qizhou and {Koch}, Patrick M. and {Kim}, Kee-Tae and {Qiu},
		  Keping and {Evans}, II, Neal J. and {Johnstone}, Doug and
		  {Thompson}, Mark and {Ward-Thompson}, Derek and {Di
		  Francesco}, James and {Tang}, Ya-Wen and {Montillaud},
		  Julien and {Kim}, Gwanjeong and {Mairs}, Steve and
		  {Sanhueza}, Patricio and {Kim}, Shinyoung and {Berry},
		  David and {Gordon}, Michael S. and {Tatematsu}, Ken'ichi
		  and {Liu}, Sheng-Yuan and {Pattle}, Kate and {Eden}, David
		  and {McGehee}, Peregrine M. and {Wang}, Ke and
		  {Ristorcelli}, I. and {Graves}, Sarah F. and {Alina}, Dana
		  and {Lacaille}, Kevin M. and {Montier}, Ludovic and {Park},
		  Geumsook and {Kwon}, Woojin and {Chung}, Eun Jung and
		  {Pelkonen}, Veli-Matti and {Micelotta}, Elisabetta R. and
		  {Saajasto}, Mika and {Fuller}, Gary},
  title		= "{Magnetic Fields in the Infrared Dark Cloud G34.43+0.24}",
  journal	= {\apj},
  year		= 2019,
  month		= sep,
  volume	= {883},
  number	= {1},
  eid		= {95},
  pages		= {95},
  doi		= {10.3847/1538-4357/ab39dd},
  archiveprefix	= {arXiv},
  eprint	= {1908.03624},
  primaryclass	= {astro-ph.GA},
  adsurl	= {https://ui.adsabs.harvard.edu/abs/2019ApJ...883...95S}
}

@Article{	  Tang_2019,
  author	= {{Tang}, Ya-Wen and {Koch}, Patrick M. and {Peretto},
		  Nicolas and {Novak}, Giles and {Duarte-Cabral}, Ana and
		  {Chapman}, Nicholas L. and {Hsieh}, Pei-Ying and {Yen},
		  Hsi-Wei},
  title		= "{Gravity, Magnetic Field, and Turbulence: Relative
		  Importance and Impact on Fragmentation in the Infrared Dark
		  Cloud G34.43+00.24}",
  journal	= {\apj},
  year		= 2019,
  month		= jun,
  volume	= {878},
  number	= {1},
  eid		= {10},
  pages		= {10},
  doi		= {10.3847/1538-4357/ab1484},
  archiveprefix	= {arXiv},
  eprint	= {1903.12397},
  primaryclass	= {astro-ph.GA},
  adsurl	= {https://ui.adsabs.harvard.edu/abs/2019ApJ...878...10T}
}

@Article{	  Lopez_2020,
  author	= {{A{\~n}ez-L{\'o}pez}, N. and {Busquet}, G. and {Koch},
		  P.~M. and {Girart}, J.~M. and {Liu}, H.~B. and {Santos}, F.
		  and {Chapman}, N.~L. and {Novak}, G. and {Palau}, A. and
		  {Ho}, P.~T.~P. and {Zhang}, Q.},
  title		= "{Role of the magnetic field in the fragmentation process:
		  the case of G14.225-0.506}",
  journal	= {\aap},
  year		= 2020,
  month		= dec,
  volume	= {644},
  eid		= {A52},
  pages		= {A52},
  doi		= {10.1051/0004-6361/202039152},
  archiveprefix	= {arXiv},
  eprint	= {2010.13503},
  primaryclass	= {astro-ph.GA},
  adsurl	= {https://ui.adsabs.harvard.edu/abs/2020A&A...644A..52A}
}

@Article{	  Santos_2016,
  author	= {{Santos}, F{\'a}bio P. and {Busquet}, Gemma and {Franco},
		  Gabriel A.~P. and {Girart}, Josep Miquel and {Zhang},
		  Qizhou},
  title		= "{Magnetically Dominated Parallel Interstellar Filaments in
		  the Infrared Dark Cloud G14.225-0.506}",
  journal	= {\apj},
  year		= 2016,
  month		= dec,
  volume	= {832},
  number	= {2},
  eid		= {186},
  pages		= {186},
  doi		= {10.3847/0004-637X/832/2/186},
  archiveprefix	= {arXiv},
  eprint	= {1609.08052},
  primaryclass	= {astro-ph.GA},
  adsurl	= {https://ui.adsabs.harvard.edu/abs/2016ApJ...832..186S}
}

@Article{	  Jadhav_2025,
  author	= {{Jadhav}, O.~R. and {Dewangan}, L.~K. and {Ismail}, A. Haj
		  and {Bhadari}, N.~K. and {Maity}, A.~K. and {Yadav}, Ram
		  Kesh and {Salouci}, Moustafa and {Sanhueza}, Patricio and
		  {Sharma}, Saurabh},
  title		= "{Unveiling Physical Conditions and Star Formation
		  Processes in the G47 Filamentary Cloud}",
  journal	= {\apj},
  year		= 2025,
  month		= jun,
  volume	= {986},
  number	= {1},
  eid		= {48},
  pages		= {48},
  doi		= {10.3847/1538-4357/adcee4},
  archiveprefix	= {arXiv},
  eprint	= {2504.04742},
  primaryclass	= {astro-ph.GA},
  adsurl	= {https://ui.adsabs.harvard.edu/abs/2025ApJ...986...48J}
}

@Article{	  Stephens_2022,
  author	= {{Stephens}, Ian W. and {Myers}, Philip C. and {Zucker},
		  Catherine and {Jackson}, James M. and {Andersson}, B.-G.
		  and {Smith}, Rowan and {Soam}, Archana and {Battersby},
		  Cara and {Sanhueza}, Patricio and {Hogge}, Taylor and
		  {Smith}, Howard A. and {Novak}, Giles and {Sadavoy}, Sarah
		  and {Pillai}, Thushara G.~S. and {Li}, Zhi-Yun and
		  {Looney}, Leslie W. and {Sugitani}, Koji and {Coud{\'e}},
		  Simon and {Guzm{\'a}n}, Andr{\'e}s and {Goodman}, Alyssa
		  and {Kusune}, Takayoshi and {Santos}, F{\'a}bio P. and
		  {Zuckerman}, Leah and {Encalada}, Frankie},
  title		= "{The Magnetic Field in the Milky Way Filamentary Bone
		  G47}",
  journal	= {\apjl},
  year		= 2022,
  month		= feb,
  volume	= {926},
  number	= {1},
  eid		= {L6},
  pages		= {L6},
  doi		= {10.3847/2041-8213/ac4d8f},
  archiveprefix	= {arXiv},
  eprint	= {2201.11933},
  primaryclass	= {astro-ph.GA},
  adsurl	= {https://ui.adsabs.harvard.edu/abs/2022ApJ...926L...6S}
}

@Article{	  Simon_2006,
  author	= {{Simon}, R. and {Rathborne}, J.~M. and {Shah}, R.~Y. and
		  {Jackson}, J.~M. and {Chambers}, E.~T.},
  title		= "{The Characterization and Galactic Distribution of
		  Infrared Dark Clouds}",
  journal	= {\apj},
  year		= 2006,
  month		= dec,
  volume	= {653},
  number	= {2},
  pages		= {1325-1335},
  doi		= {10.1086/508915},
  adsurl	= {https://ui.adsabs.harvard.edu/abs/2006ApJ...653.1325S}
}

@Article{	  Leurini_2011,
  author	= {{Leurini}, S. and {Pillai}, T. and {Stanke}, T. and
		  {Wyrowski}, F. and {Testi}, L. and {Schuller}, F. and
		  {Menten}, K.~M. and {Thorwirth}, S.},
  title		= "{The molecular distribution of the IRDC G351.77-0.51}",
  journal	= {\aap},
  year		= 2011,
  month		= sep,
  volume	= {533},
  eid		= {A85},
  pages		= {A85},
  doi		= {10.1051/0004-6361/201016380},
  archiveprefix	= {arXiv},
  eprint	= {1108.2117},
  primaryclass	= {astro-ph.GA},
  adsurl	= {https://ui.adsabs.harvard.edu/abs/2011A&A...533A..85L}
}

@Article{	  Leurini_2019,
  author	= {{Leurini}, S. and {Schisano}, E. and {Pillai}, T. and
		  {Giannetti}, A. and {Urquhart}, J. and {Csengeri}, T. and
		  {Casu}, S. and {Cunningham}, M. and {Elia}, D. and {Jones},
		  P.~A. and {K{\"o}nig}, C. and {Molinari}, S. and {Stanke},
		  T. and {Testi}, L. and {Wyrowski}, F. and {Menten}, K.~M.},
  title		= "{Characterising the high-mass star forming filament
		  G351.776-0.527 with Herschel and APEX dust continuum and
		  gas observations}",
  journal	= {\aap},
  year		= 2019,
  month		= jan,
  volume	= {621},
  eid		= {A130},
  pages		= {A130},
  doi		= {10.1051/0004-6361/201833612},
  archiveprefix	= {arXiv},
  eprint	= {1812.01035},
  primaryclass	= {astro-ph.GA},
  adsurl	= {https://ui.adsabs.harvard.edu/abs/2019A&A...621A.130L}
}

@Article{	  Ryabukhina_2021,
  author	= {{Ryabukhina}, O.~L. and {Zinchenko}, I.~I.},
  title		= "{A multiline study of the filamentary infrared dark cloud
		  G351.78-0.54}",
  journal	= {\mnras},
  year		= 2021,
  month		= jul,
  volume	= {505},
  number	= {1},
  pages		= {726-737},
  doi		= {10.1093/mnras/stab1309},
  archiveprefix	= {arXiv},
  eprint	= {2105.03133},
  primaryclass	= {astro-ph.GA},
  adsurl	= {https://ui.adsabs.harvard.edu/abs/2021MNRAS.505..726R}
}

@Article{	  Beuther_2025,
  author	= {{Beuther}, H. and {Olguin}, F.~A. and {Sanhueza}, P. and
		  {Cunningham}, N. and {Ginsburg}, A.},
  title		= "{Hierarchical accretion flow from the G351 infrared dark
		  filament to its central cores}",
  journal	= {\aap},
  year		= 2025,
  month		= mar,
  volume	= {695},
  eid		= {A51},
  pages		= {A51},
  doi		= {10.1051/0004-6361/202452754},
  archiveprefix	= {arXiv},
  eprint	= {2502.13866},
  primaryclass	= {astro-ph.GA},
  adsurl	= {https://ui.adsabs.harvard.edu/abs/2025A&A...695A..51B}
}

@Article{	  Norris_1993,
  author	= {{Norris}, R.~P. and {Whiteoak}, J.~B. and {Caswell}, J.~L.
		  and {Wieringa}, M.~H. and {Gough}, R.~G.},
  title		= "{Synthesis Images of 6.7 GHz Methanol Masers}",
  journal	= {\apj},
  year		= 1993,
  month		= jul,
  volume	= {412},
  pages		= {222},
  doi		= {10.1086/172914},
  adsurl	= {https://ui.adsabs.harvard.edu/abs/1993ApJ...412..222N}
}

@Article{	  Walsh_1998,
  author	= {{Walsh}, A.~J. and {Burton}, M.~G. and {Hyland}, A.~R. and
		  {Robinson}, G.},
  title		= "{Studies of ultracompact HII regions - II. High-resolution
		  radio continuum and methanol maser survey}",
  journal	= {\mnras},
  year		= 1998,
  month		= dec,
  volume	= {301},
  number	= {3},
  pages		= {640-698},
  doi		= {10.1046/j.1365-8711.1998.02014.x},
  adsurl	= {https://ui.adsabs.harvard.edu/abs/1998MNRAS.301..640W}
}

@Article{	  Beuther_2017,
  author	= {{Beuther}, H. and {Ahmadi}, A. and {Mottram}, J. and
		  {Bosco}, F. and {Linz}, H. and {Klaassen}, P. and {CORE
		  Team}},
  title		= "{CORE: fragmentation and disk formation in high-mass star
		  formation . An IRAM NOEMA large program}",
  journal	= {\memsai},
  year		= 2017,
  month		= jan,
  volume	= {88},
  pages		= {584},
  adsurl	= {https://ui.adsabs.harvard.edu/abs/2017MmSAI..88..584B}
}

@Article{	  Beuther_2019,
  author	= {{Beuther}, H. and {Ahmadi}, A. and {Mottram}, J.~C. and
		  {Linz}, H. and {Maud}, L.~T. and {Henning}, Th. and
		  {Kuiper}, R. and {Walsh}, A.~J. and {Johnston}, K.~G. and
		  {Longmore}, S.~N.},
  title		= "{High-mass star formation at sub-50 au scales}",
  journal	= {\aap},
  year		= 2019,
  month		= jan,
  volume	= {621},
  eid		= {A122},
  pages		= {A122},
  doi		= {10.1051/0004-6361/201834064},
  archiveprefix	= {arXiv},
  eprint	= {1811.10245},
  primaryclass	= {astro-ph.SR},
  adsurl	= {https://ui.adsabs.harvard.edu/abs/2019A&A...621A.122B}
}

@Article{	  Leurini_2008,
  author	= {{Leurini}, S. and {Hieret}, C. and {Thorwirth}, S. and
		  {Wyrowski}, F. and {Schilke}, P. and {Menten}, K.~M. and
		  {G{\"u}sten}, R. and {Zapata}, L.},
  title		= "{High-mass star formation in the IRAS 17233-3606 region: a
		  new nearby and bright hot core in the southern sky}",
  journal	= {\aap},
  year		= 2008,
  month		= jul,
  volume	= {485},
  number	= {1},
  pages		= {167-175},
  doi		= {10.1051/0004-6361:200809475},
  archiveprefix	= {arXiv},
  eprint	= {0804.4495},
  primaryclass	= {astro-ph},
  adsurl	= {https://ui.adsabs.harvard.edu/abs/2008A&A...485..167L}
}

@Article{	  Zapata_2008,
  author	= {{Zapata}, Luis A. and {Leurini}, Silvia and {Menten}, Karl
		  M. and {Schilke}, Peter and {Rolffs}, Rainer and {Hieret},
		  Carolin},
  title		= "{Unveiling a Compact Cluster of Massive and Young Stars in
		  IRAS 17233-3606}",
  journal	= {\aj},
  year		= 2008,
  month		= oct,
  volume	= {136},
  number	= {4},
  pages		= {1455-1462},
  doi		= {10.1088/0004-6256/136/4/1455},
  archiveprefix	= {arXiv},
  eprint	= {0807.1591},
  primaryclass	= {astro-ph},
  adsurl	= {https://ui.adsabs.harvard.edu/abs/2008AJ....136.1455Z}
}

@Article{	  Leurini_2009,
  author	= {{Leurini}, S. and {Codella}, C. and {Zapata}, L.~A. and
		  {Belloche}, A. and {Stanke}, T. and {Wyrowski}, F. and
		  {Schilke}, P. and {Menten}, K.~M. and {G{\"u}sten}, R.},
  title		= "{Extremely high velocity gas from the massive young
		  stellar objects in IRAS 17233-3606}",
  journal	= {\aap},
  year		= 2009,
  month		= dec,
  volume	= {507},
  number	= {3},
  pages		= {1443-1454},
  doi		= {10.1051/0004-6361/200912783},
  archiveprefix	= {arXiv},
  eprint	= {0909.0525},
  primaryclass	= {astro-ph.SR},
  adsurl	= {https://ui.adsabs.harvard.edu/abs/2009A&A...507.1443L}
}

@Article{	  leurini_2011b,
  author	= {{Leurini}, S. and {Codella}, C. and {Zapata}, L. and
		  {Beltr{\'a}n}, M.~T. and {Schilke}, P. and {Cesaroni}, R.},
  title		= "{On the kinematics of massive star forming regions: the
		  case of IRAS 17233-3606}",
  journal	= {\aap},
  year		= 2011,
  month		= jun,
  volume	= {530},
  eid		= {A12},
  pages		= {A12},
  doi		= {10.1051/0004-6361/201016190},
  archiveprefix	= {arXiv},
  eprint	= {1104.0857},
  primaryclass	= {astro-ph.SR},
  adsurl	= {https://ui.adsabs.harvard.edu/abs/2011A&A...530A..12L}
}

@Article{	  Leurini_2013,
  author	= {{Leurini}, S. and {Codella}, C. and {Gusdorf}, A. and
		  {Zapata}, L. and {G{\'o}mez-Ruiz}, A. and {Testi}, L. and
		  {Pillai}, T.},
  title		= "{Evidence of a SiO collimated outflow from a massive YSO
		  in IRAS 17233-3606}",
  journal	= {\aap},
  year		= 2013,
  month		= jun,
  volume	= {554},
  eid		= {A35},
  pages		= {A35},
  doi		= {10.1051/0004-6361/201118154},
  archiveprefix	= {arXiv},
  eprint	= {1304.4401},
  primaryclass	= {astro-ph.GA},
  adsurl	= {https://ui.adsabs.harvard.edu/abs/2013A&A...554A..35L}
}

@Article{	  Klaassen_2015,
  author	= {{Klaassen}, P.~D. and {Johnston}, K.~G. and {Leurini}, S.
		  and {Zapata}, L.~A.},
  title		= "{The SiO outflow from IRAS 17233-3606 at high resolution}",
  journal	= {\aap},
  year		= 2015,
  month		= mar,
  volume	= {575},
  eid		= {A54},
  pages		= {A54},
  doi		= {10.1051/0004-6361/201424781},
  archiveprefix	= {arXiv},
  eprint	= {1412.5823},
  primaryclass	= {astro-ph.SR},
  adsurl	= {https://ui.adsabs.harvard.edu/abs/2015A&A...575A..54K}
}

@Article{	  Motte_2022,
  author	= {{Motte}, F. and {Bontemps}, S. and {Csengeri}, T. and
		  {Pouteau}, Y. and {Louvet}, F. and {Stutz}, A.~M. and
		  {Cunningham}, N. and {L{\'o}pez-Sepulcre}, A. and
		  {Brouillet}, N. and {Galv{\'a}n-Madrid}, R. and {Ginsburg},
		  A. and {Maud}, L. and {Men'shchikov}, A. and {Nakamura}, F.
		  and {Nony}, T. and {Sanhueza}, P. and
		  {{\'A}lvarez-Guti{\'e}rrez}, R.~H. and {Armante}, M. and
		  {Baug}, T. and {Bonfand}, M. and {Busquet}, G. and
		  {Chapillon}, E. and {D{\'\i}az-Gonz{\'a}lez}, D. and
		  {Fern{\'a}ndez-L{\'o}pez}, M. and {Guzm{\'a}n}, A.~E. and
		  {Herpin}, F. and {Liu}, H.-L. and {Olguin}, F. and
		  {Towner}, A.~P.~M. and {Bally}, J. and {Battersby}, C. and
		  {Braine}, J. and {Bronfman}, L. and {Chen}, H.-R.~V. and
		  {Dell'Ova}, P. and {Di Francesco}, J. and {Gonz{\'a}lez},
		  M. and {Gusdorf}, A. and {Hennebelle}, P. and {Izumi}, N.
		  and {Joncour}, I. and {Lee}, Y.-N. and {Lefloch}, B. and
		  {Lesaffre}, P. and {Lu}, X. and {Menten}, K.~M. and
		  {Mignon-Risse}, R. and {Molet}, J. and {Moraux}, E. and
		  {Mundy}, L. and {Nguyen Luong}, Q. and {Reyes}, N. and
		  {Reyes Reyes}, S.~D. and {Robitaille}, J.-F. and
		  {Rosolowsky}, E. and {Sandoval-Garrido}, N.~A. and
		  {Schuller}, F. and {Svoboda}, B. and {Tatematsu}, K. and
		  {Thomasson}, B. and {Walker}, D. and {Wu}, B. and
		  {Whitworth}, A.~P. and {Wyrowski}, F.},
  title		= "{ALMA-IMF. I. Investigating the origin of stellar masses:
		  Introduction to the Large Program and first results}",
  journal	= {\aap},
  year		= 2022,
  month		= jun,
  volume	= {662},
  eid		= {A8},
  pages		= {A8},
  doi		= {10.1051/0004-6361/202141677},
  archiveprefix	= {arXiv},
  eprint	= {2112.08182},
  primaryclass	= {astro-ph.GA},
  adsurl	= {https://ui.adsabs.harvard.edu/abs/2022A&A...662A...8M}
}

@Article{	  Reyes_2024,
  author	= {{Reyes-Reyes}, S.~D. and {Stutz}, A.~M. and {Megeath},
		  S.~T. and {Xu}, Fengwei and {{\'A}lvarez-Guti{\'e}rrez},
		  R.~H. and {Sandoval-Garrido}, N. and {Liu}, H.-L.},
  title		= "{Benchmarking the IRDC G351.77-0.53: Gaia DR3 distance,
		  mass distribution, and star formation content}",
  journal	= {\mnras},
  year		= 2024,
  month		= apr,
  volume	= {529},
  number	= {3},
  pages		= {2220-2233},
  doi		= {10.1093/mnras/stae631},
  archiveprefix	= {arXiv},
  eprint	= {2403.02456},
  primaryclass	= {astro-ph.GA},
  adsurl	= {https://ui.adsabs.harvard.edu/abs/2024MNRAS.529.2220R}
}

@Article{	  Benjamin_2003,
  doi		= {10.1086/376696},
  url		= {https://dx.doi.org/10.1086/376696},
  year		= {2003},
  month		= {aug},
  publisher	= {The University of Chicago Press},
  volume	= {115},
  number	= {810},
  pages		= {953},
  author	= {Robert A. Benjamin and E. Churchwell and Brian L. Babler
		  and T. M. Bania and Dan P. Clemens and Martin Cohen and
		  John M. Dickey and Rémy Indebetouw and James M. Jackson
		  and Henry A. Kobulnicky and Alex Lazarian and A. P. Marston
		  and John S. Mathis and Marilyn R. Meade and Sara Seager and
		  S. R. Stolovy and C. Watson and Barbara A. Whitney and
		  Michael J. Wolff and Mark G. Wolfire},
  title		= {GLIMPSE. I. An SIRTF Legacy Project to Map the Inner
		  Galaxy},
  journal	= {Publications of the Astronomical Society of the Pacific}
}

@Article{	  Lang_2014,
  author	= {{Lang}, Dustin},
  title		= "{unWISE: Unblurred Coadds of the WISE Imaging}",
  journal	= {\aj},
  year		= 2014,
  month		= may,
  volume	= {147},
  number	= {5},
  eid		= {108},
  pages		= {108},
  doi		= {10.1088/0004-6256/147/5/108},
  archiveprefix	= {arXiv},
  eprint	= {1405.0308},
  primaryclass	= {astro-ph.IM},
  adsurl	= {https://ui.adsabs.harvard.edu/abs/2014AJ....147..108L}
}

@Article{	  Urquhart_2017,
  author	= {Urquhart, J. S. and König, C. and Giannetti, A. and
		  Leurini, S. and Moore, T. J. T. and Eden, D. J. and Pillai,
		  T. and Thompson, M. A. and Braiding, C. and Burton, M. G.
		  and Csengeri, T. and Dempsey, J. T. and Figura, C. and
		  Froebrich, D. and Menten, K. M. and Schuller, F. and Smith,
		  M. D. and Wyrowski, F.},
  title		= "{ATLASGAL – properties of a complete sample of Galactic
		  clumps}",
  journal	= {Monthly Notices of the Royal Astronomical Society},
  volume	= {473},
  number	= {1},
  pages		= {1059-1102},
  year		= {2017},
  month		= {09},
  issn		= {0035-8711},
  doi		= {10.1093/mnras/stx2258},
  url		= {https://doi.org/10.1093/mnras/stx2258},
  eprint	= {https://academic.oup.com/mnras/article-pdf/473/1/1059/21408259/stx2258.pdf}
}

@Article{	  Harper_2018,
  author	= {Harper, Doyal A. and Runyan, Marcus C. and Dowell, C.
		  Darren and Wirth, C. Jesse and Amato, Michael and Ames,
		  Troy and Amiri, Mandana and Banks, Stuart and Bartels,
		  Arlin and Benford, Dominic J. and Berthoud, Marc and
		  Buchanan, Ernest and Casey, Sean and Chapman, Nicholas L.
		  and Chuss, David T. and Cook, Brant and Derro, Rebecca and
		  Dotson, Jessie L. and Evans, Rhodri and Fixsen, Dale and
		  Gatley, Ian and Guerra, Jordan A. and Halpern, Mark and
		  Hamilton, Ryan T. and Hamlin, Louise A. and Hansen,
		  Christopher J. and Heimsath, Stephen and Hermida, Alfonso
		  and Hilton, Gene C. and Hirsch, Robert and Hollister,
		  Matthew I. and Hostetter, Carl F. and Irwin, Kent and
		  Jhabvala, Christine A. and Jhabvala, Murzban and Kastner,
		  Joel and Kov\'{a}cs, Attila and Lin, Sean and Loewenstein,
		  Robert F. and Looney, Leslie W. and Lopez-Rodriguez,
		  Enrique and Maher, Stephen F. and Michail, Joseph M. and
		  Miller, Timothy M. and Moseley, S. Harvey and Novak, Giles
		  and Pernic, Robert J. and Rennick, Timothy and Rhody,
		  Harvey and Sandberg, Eric and Sandford, Dale and Santos,
		  Fabio Pereira and Shafer, Rick and Sharp, Elmer H. and
		  Shirron, Peter and Siah, Javad and Silverberg, Robert and
		  Sparr, Leroy M. and Spotz, Robert and Staguhn, Johannes G.
		  and Toorian, Armen S. and Towey, Shannon and Tuttle, Jim
		  and Vaillancourt, John and Voellmer, George and Volpert,
		  Carolyn G. and Wang, Shu-i and Wollack, Edward J.},
  title		= {HAWC+, the Far-Infrared Camera and Polarimeter for SOFIA},
  journal	= {Journal of Astronomical Instrumentation},
  volume	= {07},
  number	= {04},
  pages		= {1840008},
  year		= {2018},
  doi		= {10.1142/S2251171718400081},
  url		= {
		  
		  https://doi.org/10.1142/S2251171718400081
		  
		  },
  eprint	= {
		  
		  https://doi.org/10.1142/S2251171718400081
		  
		  }
}

@Article{	  Schuller_2009,
  author	= {{Schuller, F.} and {Menten, K. M.} and {Contreras, Y.} and
		  {Wyrowski, F.} and {Schilke, P.} and {Bronfman, L.} and
		  {Henning, T.} and {Walmsley, C. M.} and {Beuther, H.} and
		  {Bontemps, S.} and {Cesaroni, R.} and {Deharveng, L.} and
		  {Garay, G.} and {Herpin, F.} and {Lefloch, B.} and {Linz,
		  H.} and {Mardones, D.} and {Minier, V.} and {Molinari, S.}
		  and {Motte, F.} and {Nyman, L.-Å.} and {Reveret, V.} and
		  {Risacher, C.} and {Russeil, D.} and {Schneider, N.} and
		  {Testi, L.} and {Troost, T.} and {Vasyunina, T.} and
		  {Wienen, M.} and {Zavagno, A.} and {Kovacs, A.} and
		  {Kreysa, E.} and {Siringo, G.} and {Weiß, A.}},
  title		= "{ATLASGAL – The APEX telescope large area survey of the
		  galaxy at 870 $\mathsf{\mu}$m}",
  doi		= "10.1051/0004-6361/200811568",
  url		= "https://doi.org/10.1051/0004-6361/200811568",
  journal	= {Astronomy and Astrophysics Journal},
  year		= 2009,
  volume	= 504,
  number	= 2,
  pages		= "415-427"
}

@Article{	  Mouschovias_1976,
  author	= {{Mouschovias}, T. Ch. and {Spitzer}, Jr., L.},
  title		= "{Note on the collapse of magnetic interstellar clouds.}",
  journal	= {\apj},
  year		= 1976,
  month		= dec,
  volume	= {210},
  pages		= {326},
  doi		= {10.1086/154835},
  adsurl	= {https://ui.adsabs.harvard.edu/abs/1976ApJ...210..326M}
}

@Article{	  Gomez_2018,
  author	= {{G{\'o}mez}, Gilberto C. and {V{\'a}zquez-Semadeni},
		  Enrique and {Zamora-Avil{\'e}s}, Manuel},
  title		= "{The magnetic field structure in molecular cloud
		  filaments}",
  journal	= {\mnras},
  year		= 2018,
  month		= nov,
  volume	= {480},
  number	= {3},
  pages		= {2939-2944},
  doi		= {10.1093/mnras/sty2018},
  archiveprefix	= {arXiv},
  eprint	= {1801.03169},
  primaryclass	= {astro-ph.GA},
  adsurl	= {https://ui.adsabs.harvard.edu/abs/2018MNRAS.480.2939G}
}

\end{document}